\documentclass[english,aps,preprint,nofootinbib,superscriptaddress]{revtex4}
\pdfoutput=1
\usepackage[T1]{fontenc}
\usepackage[latin9]{inputenc}
\usepackage{color}
\definecolor{note_fontcolor}{rgb}{0.80078125, 0.80078125, 0.80078125}
\usepackage{verbatim}
\usepackage{amsmath}
\usepackage{graphicx}
\usepackage[colorlinks=true,linkcolor=blue,citecolor=blue,filecolor=blue,urlcolor=blue]{hyperref}
\makeatletter

\@ifundefined{textcolor}{}
{%
 \definecolor{BLACK}{gray}{0}
 \definecolor{WHITE}{gray}{1}
 \definecolor{RED}{rgb}{1,0,0}
 \definecolor{GREEN}{rgb}{0,1,0}
 \definecolor{BLUE}{rgb}{0,0,1}
 \definecolor{CYAN}{cmyk}{1,0,0,0}
 \definecolor{MAGENTA}{cmyk}{0,1,0,0}
 \definecolor{YELLOW}{cmyk}{0,0,1,0}
 }

\usepackage{bbm}

\makeatother

\usepackage{babel}
\begin{document}
\global\long\def\abs#1{\left| #1 \right| }
\global\long\def\half{\frac{1}{2}}
\global\long\def\partder#1#2{\frac{\partial#1}{\partial#2}}
 \global\long\def\comm#1#2{\left[ #1 ,#2 \right] }

\global\long\def\Tr#1{\textrm{Tr}\left\{  #1 \right\}  }

\global\long\def\Imag#1{\mathrm{Im}\left\{  #1 \right\}  }

\global\long\def\Real#1{\mathrm{Re}\left\{  #1 \right\}  }

\global\long\def\db{\!\not\!\! D\,}

\global\long\def\gesim{\,{\raise-3pt\hbox{$\sim$}}\!\!\!\!\!{\raise2pt\hbox{$>$}}\,}
\global\long\def\then{{\quad\Rightarrow\quad}}
\global\long\def\lcal{{\cal L}}
\global\long\def\mcal{{\cal M}}
\global\long\def\bBB{{\mathbbm B}}
\global\long\def\sigbf{\bm{\sigma}}
\global\long\def\gev{\hbox{GeV}}
\global\long\def\tev{\hbox{TeV}}
\global\long\def\vevof#1{\left\langle #1\right\rangle }
\global\long\def\up#1{^{\left( #1 \right) }}
\global\long\def\inv#1{\frac{1}{#1}}
\global\long\def\su#1{{SU(#1)}}
\global\long\def\ui{U(1)}

\preprint{CAFPE-165/11, UG-FT-295/11, FTUV-11-1128, IFIC/11-65, UCRHEP-T512}

\title{A realistic model of neutrino masses with a large neutrinoless double beta decay rate}

\author{Francisco~del~Aguila}
\affiliation{CAFPE and Departamento de Fisica Teorica y del Cosmos, Universidad
de Granada, E\textendash{}18071 Granada, Spain}

\author{Alberto~Aparici}
\affiliation{Departament de Fisica Teorica, Universitat de Valencia and IFIC,
Universitat de Valencia-CSIC, Dr. Moliner 50, E-46100 Burjassot (Valencia),
Spain}

\author{Subhaditya~Bhattacharya}
\affiliation{Department of Physics and Astronomy, University of California, Riverside
CA 92521-0413, USA}

\author{Arcadi~Santamaria}
\affiliation{Departament de Fisica Teorica, Universitat de Valencia and IFIC,
Universitat de Valencia-CSIC, Dr. Moliner 50, E-46100 Burjassot (Valencia),
Spain}

\author{Jose~Wudka}
\affiliation{Department of Physics and Astronomy, University of California, Riverside CA 92521-0413, USA}

\begin{abstract}
The minimal Standard Model extension with the Weinberg operator 
does accommodate the observed neutrino masses and mixing, 
but predicts a neutrinoless double beta ($0\nu\beta\beta$) decay rate proportional 
to the effective electron neutrino mass, which can be then arbitrarily small 
within present experimental limits.
However, in general $0\nu\beta\beta$ decay can have an independent origin and 
be near its present experimental bound; whereas neutrino masses are generated 
radiatively, contributing negligibly to $0\nu\beta\beta$ decay. 
We provide a realization of this scenario in a simple, well
defined and testable model, with potential LHC effects
and calculable neutrino masses, whose two-loop expression
we derive exactly. We also discuss the connection of
this model to others that have appeared in the literature,
and remark on the significant differences that result
from various choices of quantum number assignments and symmetry
assumptions.
In this type of models lepton flavor violating rates are also preferred to be 
relatively large, at the reach of foreseen experiments. 
Interestingly enough, in our model this stands for a large third mixing angle, 
$\sin^2\theta_{13} \gtrsim 0.008$, when $\mu \rightarrow eee$ is required to lie 
below its present experimental limit. 
\end{abstract}

\maketitle

\section{Introduction\label{sec:Introduction}}

Neutrino oscillations are the only new physics (NP) 
beyond the minimal Standard Model (SM) observed 
to date \cite{Nakamura:2010zzi} (for recent reviews see for instance 
\cite{GonzalezGarcia:2007ib,Mohapatra:2005wg,Altarelli:1999gu}). 
They can be fully explained introducing rather small neutrino 
masses $m_\nu$ and the corresponding (unitary) charged current 
mixing matrix $U$ \cite{Pontecorvo:1967fh,Maki:1962mu}.  
The observed pattern of neutrino masses can be implemented, 
however, in two quite different ways depending on the Dirac or 
Majorana character of the neutrinos. In the first
case the SM is extended 
by adding three SM singlets, $\nu_{R}$, to provide Dirac masses 
to the SM neutrinos, $\nu_{L}$, through small Yukawa couplings. 
Alternatively, we can consider extending the SM by adding
 the only invariant dimension 5 
(Weinberg) operator that can be written using the SM 
field content \cite{Weinberg:1979sa}~\footnote{$\phi$ and $\ell_L$ are the SM Higgs and 
lepton doublets and $\tilde\phi=i\tau_2\phi^*$,  
$\tilde{\ell}_L=i\tau_2\ell_L^c$.},
${\cal O}_5 = \overline{\tilde\ell_{L}}\phi {\tilde\phi}^\dagger \ell_{L}$. 
In this case the SM 
neutrinos acquire Majorana masses after electroweak symmetry breaking, 
$\langle \phi^0 \rangle = v_\phi$,
and are inversely proportional to $\Lambda$, the NP
scale associated with $ {\cal O}_5$. The small neutrino 
masses then require that the coefficients of this operator be small, either due 
to a very large NP scale or  suppressed 
dimensionless couplings.

Both alternatives (that light neutrinos are Dirac or 
Majorana), are viable and indistinguishable if $ \Lambda $ is  very large, 
{\em except} for the possible observation of lepton number violation (LNV) 
in neutrinoless double beta ($0\nu\beta\beta$) decay \cite{Furry:1939qr}~
\footnote{We shall not discuss the implications of requiring enough leptogenesis 
to account for the observed baryogenesis~\cite{Fukugita:1986hr} (for recent reviews 
see~\cite{Davidson:2008bu,Branco:2011zb}).} (for a recent review see~\cite{Vergados:2002pv}). 
Indeed, the SM extension with Dirac neutrino masses preserves lepton 
number (LN), and hence $0\nu\beta\beta$ decay is forbidden. 
Whereas Majorana masses carry LN equal two, as does the 
Weinberg operator, and this leads to a non-zero $0\nu\beta\beta$ width.
Thus, the observation of $0\nu\beta\beta$ decay would strongly favor Majorana 
neutrino masses~\cite{Schechter:1981bd}; hence the  prime relevance of this type of experiments 
(see \cite{Barabash:2011fg,Avignone:2007fu} for recent reviews). 

All this, however, relies on the assumption that both such minimal 
SM extensions describe the dominating NP effects up to very high scales. 
We will argue, however, that LNV and neutrino masses 
may be due to NP near the electroweak scale, in which case a much richer
set of possibilities can be realized. The main point we wish to emphasize is that
although both $0\nu\beta\beta$ decay and 
Majorana neutrino masses do violate LN, they need not be both directly 
related to the Weinberg operator ${\cal O}_5$, 
unlike the above minimal SM extension consisting only of this dimension 5 operator. 
For instance, the leading LNV effects in 
$0\nu\beta\beta$ decay could be mediated by an 
operator involving only {\em right-handed} (RH) electrons, such as, for example,
${\cal O}_9 = \left(\overline{e_{R}}e_{R }^{c}\right)
\left(\phi^{\dagger}D^{\mu}\tilde{\phi}\right)
\left(\phi^{\dagger}D_{\mu}\tilde{\phi}\right)$
which is the lowest-order LNV operator with two RH leptons 
and invariant under the SM gauge transformations.
In this case ${\cal O}_5 $ (with the associated Majorana masses) is generated radiatively 
by $ {\cal O}_9 $ at two loops, 
where the charged leptons suffer a chirality change through mass insertions and
are transformed into neutrinos through the exchange of a $W$ boson. This
results in a further suppression by two charged lepton masses divided 
by two powers 
of the NP mass scale, which we do assume to be close to the 
electroweak scale. 
These radiatively-generated Majorana masses will produce the usual contribution 
to $0\nu\beta\beta$ decay, which, however, will be negligible 
compared to the ${\cal O}_9$ one.  

In a companion paper \cite{delAguila:2011zz} we classify the different ways of  
generating  $0\nu\beta\beta$ decay and light neutrino masses 
by the addition of higher order effective operators. This has 
been studied in the literature~\cite{Babu:2001ex,Choi:2002bb,Engel:2003yr,deGouvea:2007xp}, 
but mostly for operators involving fermions and scalars; we will concentrate instead
on operators involving
gauge bosons but not quarks 
(thus, for example, excluding {\em ab initio} models with 
heavy leptoquarks from our analysis). Here we will
instead provide a realistic testable model realizing the above scenario,
where {\it(i)} LN is broken at the electroweak scale; {\it(ii)}  
$0\nu\beta\beta$ decay 
into two RH electrons has a rate of the order of its experimental 
limit, through the tree-level exchange of new scalars; and {\it(iii)} 
it contains  finite, and therefore calculable, neutrino masses.
Despite a relatively small number of parameters
this model can also accommodate the observed pattern of neutrino 
masses and mixings, which are generated at two-loop order (and whose
contribution to $0\nu\beta\beta$ decay is in this case negligible).
This model is related to others that have been discussed in the
literature \cite{Chen:2006vn,Chen:2007dc}, the differences are
crucial, and essential in maintaining the 3 features just mentioned;
we discuss these points in detail below.

There are many SM extensions where $0\nu\beta\beta$ decay 
receive new contributions besides those from light Majorana neutrino masses
(see \cite{Mohapatra:1998rq} for a general overview). 
The simplest scenario assumes the presence of heavy
Majorana neutrinos whose exchange generates 
new contributions to $0\nu\beta\beta$ decay, similar to
the ones generated by the light neutrinos.
(See for recent work \cite{Atre:2009rg,Ibarra:2010xw,Mitra:2011qr,Blennow:2010th}.) 
Other extensions with many more new particles
have also been studied; such as
left-right (LR) models \cite{Pati:1974yy,Mohapatra:1974gc,Senjanovic:1975rk} 
and supersymmetric extensions (see, for instance, for a review \cite{Nakamura:2010zzi}, and 
references there in).
In such models  there are several contributions
to the $0\nu\beta\beta $ decay amplitude, any of which can dominate
depending on the region of parameter space being considered. 
We are interested, however, in identifying the minimal SM 
extensions leading to the largest possible contributions to $0\nu\beta\beta$ decay 
and containing no  independent neutrino masses. 
This means that, as argued above,
light neutrino masses only result from the unavoidable contributions 
mediated by the LNV operators generating
$0\nu\beta\beta$ 
decay; so that the
effective Lagrangian approach provides the proper
language for classifying such scenarios~\cite{delAguila:2011zz}. 
Here, we are only concerned with giving a specific example 
of the case 
where $0\nu\beta\beta$ and neutrino masses are
generated by the same underlying physics but
at different orders in perturbation theory,
with the former appearing at tree level, while the latter
only at two loops 
\footnote{LNV effective operators including quarks also generate neutrino masses
but in general at higher loop order, in fact too high in some cases to explain the observed
spectrum~\cite{Duerr:2011zd}. Although such operators
are not considered in our analysis, we will comment on them further when
discussing our general set up
in detail~\cite{delAguila:2011zz}.}. 
Simple models with finite neutrino masses at two loops have 
been often discussed in the literature \cite{Zee:1985id,Babu:1988ki}, 
although not necessarily related to NP inducing $0\nu\beta\beta$ decay 
as it is the case here.
   
It is worth emphasizing again that the model we study is 
one particularly simple example of a class of models realizing the above 
scenario, and that all such models share many phenomenological implications. 
In our case the model contains a discrete $Z_2$ symmetry,
which is spontaneously broken. This model has the virtue of providing a direct analysis 
of the symmetries and scales, however, it also has a domain-wall 
problem~\cite{Zeldovich:1974uw,Vilenkin:1984ib}.
One way of dealing with this problem is to allow the discrete symmetry to remain unbroken, 
but at the price of generating both neutrino masses and the $0\nu\beta\beta$ amplitude at 
a higher loop order; a possibility we will not pursue. We instead discuss related but somewhat 
more involved models that avoid the domain wall problem  while retaining the same low-energy 
phenomenology. 
We will restrict ourselves to the most relevant
region of parameter space where the $0\nu\beta\beta$ decay rate lies within the 
reach of the next round of experiments. With this assumption, together 
with the  constraints from lepton flavor violation (LFV) 
processes such as $\mu \rightarrow eee$, and
with the requirement of perturbative unitarity (or if preferred, naturalness of 
perturbation theory), the model predicts that the neutrino masses 
obey a normal hierarchy, that the lightest neutrino mass lies
in the range $0.002\ {\rm eV} \lesssim m_1 \lesssim 0.007\ {\rm eV}$, 
and that the third mixing angle 
in the Pontecorvo-Maki-Nakagawa-Sakata mixing matrix $U$ 
\cite{Pontecorvo:1967fh,Maki:1962mu} 
obeys $\sin^2\theta_{13} \gtrsim 0.008$.
Value which lies well within the sensitivity of ongoing 
neutrino oscillation experiments. 
Besides, the new (charged) scalar masses 
can be within the LHC reach, and various LFV processes
are predicted to have rates that will be probed when present precision is 
improved by the next generation of experiments.

In next section we present our model; few 
details on the scalar spectrum and couplings are also worked 
out, and its relation to other models is briefly discussed. 
The reader mainly interested in the phenomenological 
implications of the model can go directly to Section \ref{sec:0nu2beta}, 
where we evaluate the rate for $0\nu\beta\beta$ 
decay. The requirement that this process can be observed in the 
next round of experiments, together with perturbativity, translates into 
upper bounds on the masses of the extra scalars. 
These, in turn, result in limits on their couplings if 
present bounds on LFV processes are to be fulfilled, as 
we show in Section IV. 
We calculate the neutrino masses in Section \ref{sec:the neutrino mass}, 
where we show that the model can 
accommodate the observed pattern of neutrino masses and mixings, 
though it predicts a somewhat small electron neutrino 
effective mass $m_{ee}$. 
Finally, the prospects for the discovery of the extra scalars at LHC 
are considered in Section \ref{sec:colliders}.
In connection with this it must be noted that models with extra scalars may allow for 
a would-be Higgs boson more elusive to LHC searches, for it can have further 
decay channels open~(see for instance~\cite{Shrock:1982kd,Bertolini:1988ma,Aparici:2009fh}). 
Conclusions are drawn in last section. 
In three Appendices we collect some technical details. 

\section{A model with lepton number softly broken
\label{sec:The-model}}

This model only extends the SM Higgs sector, in order to 
allow for scalar couplings to lepton bilinears with non-zero LN. 
More precisely, as we look for separating the origin of neutrino masses 
from the mechanism mediating $0\nu\beta\beta$ decay, we 
only introduce new scalar couplings to RH charged leptons, 
ensuring they are the only final state produced in 
tree-level $0\nu\beta\beta$ decay.  
A simple way to achieve this is to enlarge 
the Higgs sector including, besides the 
SM scalar doublet $\phi$ of hypercharge 1/2, a complex 
scalar singlet $\kappa$ of hypercharge 2, a real (neutral) scalar 
singlet $\sigma$, and an electroweak triplet $\chi$ of hypercharge 1. 
We also impose a $Z_{2}$ symmetry, under which all SM particles
and $\kappa$ are even while both $\sigma$ and $\chi$ are odd, 
which forbids their coupling to lepton bilinears. 
Thus, the Yukawa Lagrangian reads 
\begin{equation}
\mathcal{L}_{\mathrm{Y}}= \overline{\ell_L}Y_{e}e_R\,\phi+
\overline{e^{c}_R}ge_R\kappa+\mathrm{h.c.}\ ,
\label{Yuk}
\end{equation}
where we can assume $Y_{e}$ to be a $3\times 3$ diagonal matrix with 
positive eigenvalues, and $g$ 
a complex symmetric matrix with three of its phases unphysical.

The most general Higgs potential consistent with the symmetries
is~\footnote{Terms like $\Tr{\chi^{2}}\Tr{\chi^{\dagger2}}$, $\Tr{\chi^{2}\chi^{\dagger2}}$
or $\phi^{\dagger}\chi\chi^{\dagger}\phi$ can easily be related to
the terms already included by using the fact that for two arbitrary $2\times2$
traceless matrices $A=A_{i}\tau_{i}$ and $B=B_{i}\tau_{i}$, $\{A,B\}=\Tr{AB}\mathbbm{1}$. 
Other possible terms are forbidden by the discrete symmetry.}
\begin{eqnarray}
V & = & -m_{\phi}^{2}|\phi|^{2}-m_{\sigma}^{2}\sigma^{2}+m_{\chi}^{2}\Tr{\chi^{\dagger}\chi}+
m_{\kappa}^{2}|\kappa|^{2}\nonumber \\
 &  & + \lambda_{\phi}|\phi|^{4}+\lambda_{\sigma}\sigma^{4}+\lambda_{\kappa}|\kappa|^{4}+
 \lambda_{\chi}\left(\Tr{\chi^{\dagger}\chi}\right)^{2}+
 \lambda_{\chi}^{\prime}\Tr{\left(\chi^{\dagger}\chi\right)^{2}}\nonumber \\
 &  & + \lambda_{\phi\sigma}|\phi|^{2}\sigma^{2}+\lambda_{\phi\kappa}|\phi|^{2}|\kappa|^{2}+
 \lambda_{\phi\chi}|\phi|^{2}\Tr{\chi^{\dagger}\chi}+
 \lambda_{\phi\chi}^{\prime}\phi^{\dagger}\chi^{\dagger}\chi\phi\nonumber \\
 &  & + \lambda_{\sigma\kappa}\sigma^{2}|\kappa|^{2}+
 \lambda_{\kappa\chi}|\kappa|^{2}\Tr{\chi^{\dagger}\chi}+
 \lambda_{\sigma\chi}\sigma^{2}\Tr{\chi^{\dagger}\chi}\nonumber \\
 &  & + \left[\mu_{\kappa}\kappa\Tr{\left(\chi^{\dagger}\right)^{2}}+
 \lambda_{6}\sigma\phi^{\dagger}\chi\tilde{\phi}+\mathrm{h.c.}\right] 
 \ ,
\label{eq:model.L}
\end{eqnarray}
where by rephasing the fields we can always choose $\mu_{\kappa}$
and $\lambda_{6}$ real of either sign. For convenience, 
we will take $\mu_{\kappa}$ positive and $\lambda_{6}$ negative. In terms
of charge eigenstates, the fields $\chi$ and $\phi$ are written 
\begin{equation}
\chi=\begin{pmatrix}\chi^{+}/\sqrt{2} & \chi^{++}\\
\chi^{0} & -\chi^{+}/\sqrt{2}
\end{pmatrix}\,,\qquad\phi=\begin{pmatrix}\phi^{+}\\
\phi^{0}
\end{pmatrix} \ .
\end{equation}
The singlet $\sigma$ is introduced to preserve the discrete symmetry that forbids 
the scalar triplet from coupling to leptons. This
symmetry is  broken spontaneously by the VEVs $ \langle\sigma\rangle$ and
$\langle\chi^0\rangle$. 

For the phenomenological discussion it is important to note 
that in the limit of vanishing $Y_{e}$, lepton number can be defined 
to only act on $\ell_L$; this is enough to protect neutrinos from getting 
a Majorana mass. While the $e_R$ lepton number, which would forbid  
$0\nu\beta\beta$ decay (into two $e_{R}$), is explicitly broken 
in the scalar potential due to the presence of the  $\mu_\kappa$  and 
$\lambda_6$ terms.
Therefore, the $0\nu\beta\beta$ decay amplitude will be proportional to all three couplings: 
$g_{ab}$ in Eq. \eqref{Yuk} and $\mu_{\kappa}$ and $\lambda_{6}$ 
in Eq.\eqref{eq:model.L}; whereas neutrino masses will also depend on $Y_{e}$.

This is one of a class of models with the same low energy physics. 
In order to understand the common LNV features it is convenient
to consider the above theory, but  with $\sigma $ complex,
and the modifications needed to insure a 
real Lagrangian. 
In this case LN is exactly conserved; if leptons are assigned LN $-1$,
$\kappa$ carries LN equal to $2$, while 
$\sigma^\dagger$ and $\chi $ both carry LN equal to $1$,
so that type II see-saw Yukawa couplings 
$\overline{\tilde{\ell}_L} \chi \ell_L$ are forbidden.
The vertex $\sigma^\dagger \overline{\tilde{\ell}_L} \chi \ell_L$, 
however, is allowed and generated at two loops,
providing light neutrinos a finite mass
(which we calculate exactly) 
after spontaneous symmetry breaking proportional to 
$\vevof{ \sigma ^\dagger \chi }$. 
Obviously, this model has a Majoron~\cite{Chikashige:1980ui} because LN, which is a global abelian 
symmetry, is spontaneously broken. One can 
then consider the region of parameter space where
such interesting models are phenomenologically viable~\footnote{
In this model the Majoron will be mainly a singlet~\cite{Chikashige:1980ui}.
In which case its couplings to ordinary matter are small and then little constrained, while its
coupling to the Higgs boson 
is essentially free. Thus, singlet Majorons can result in invisible Higgs decays 
(see \cite{Ghosh:2011qc} for a recent example). Moreover, they can have interesting implications
in astrophysics and cosmology because they can substantially affect the cooling of supernovas~\cite{Choi:1989hi} and the
neutrino relic abundance, due to the possibility of neutrino decaying~\cite{Chikashige:1980qk} or annihilating~\cite{Choi:1991aa}
into Majorons. They can be even used as a dark matter candidate (see for instance~\cite{Gu:2010ys}) if massive (pseudo-Majorons).}, 
or promote the symmetry to 
a local one by gauging, for instance, baryon minus lepton number (B$-$L). 
Although, this is an interesting possibility, too, with renewed experimental 
interest (see, for instance, for a review \cite{Basso:2011hn}),  
it requires adding RH neutrinos 
that provide
a new potential source of 
light neutrino masses; this lies outside the goal
of the present investigation that is centered on theories where
neutrino masses are generated by the LNV effective operator accounting 
for $0\nu\beta\beta$ decay. 

For this reason we have assumed $\sigma$ to be real, 
breaking LN explicitly but leaving as a remnant an exact $Z_2$ 
symmetry. 
The general features of the theory
remain, though the Majoron does not appear and LN is reduced
to the above $Z_2$ symmetry~\footnote{One can also assign an odd $Z_2$ parity to the leptons, 
in which case this discrete symmetry equals $(-1)^{\rm LN}$.}.
The spontaneous breaking of
this discrete symmetry will generate domain walls. This poses no
problem provided $\vevof\sigma$ is sufficiently large (with
$ \lambda_6 $ correspondingly small while keeping
their product fixed)
guaranteeing the formation of such defects before the
inflationary epoch. 
We have assumed
instead that $\vevof\sigma $ is not too large because we prefer 
avoid requiring $\lambda_6$  unnaturally small;
we may then avoid a domain-wall problem by adding soft-breaking terms,
such as $\sigma^3 $ to the potential (such a modification would not require the introduction
of tree-level Yukawa interactions violating the discrete $Z_2$
symmetry). We do not pursue this discussion because within this class 
of models there is an even simpler one, with the same 
neutrino physics at low energy, and none of these 
potential drawbacks. 
It is the same model presented above, but with $\sigma $
replaced by its vacuum expectation value, $ \sigma \to
\vevof\sigma $; up to coupling constant redefinition this yields
the same potential as Eq. (\ref{eq:model.L}) without the
terms containing $\sigma $, {\em except} for the last term
that becomes $\mu_\chi \phi^{\dagger}\chi\tilde{\phi} $,
with
\begin{equation}
\mu_\chi = \lambda_6 \vevof\sigma \, , 
\label{eq:limit}
\end{equation}
of the order of a TeV.
Obviously, the resulting renormalizable Lagrangian has the same quantum 
behaviour than ours; in particular,
neutrino masses are finite and generated at two loop order
and can be obtained from our results by eliminating $\lambda_6$
using (\ref{eq:limit}). 
We will 
prefer to discuss the model including the real scalar field $\sigma$ 
because the perturbative and symmetry analyses appear to be
more transparent to us. 
In the minimal model with the SM addition of only $\kappa$ and 
$\chi$, and the scalar potential terms relating their LN charges 
$\mu_{\kappa}\kappa\Tr{\left(\chi^{\dagger}\right)^{2}}$ and 
$\mu_\chi \phi^{\dagger}\chi\tilde{\phi}$ the LN of the $\chi$ field is not 
well defined since the first term requires it to be 1 while the second 0 (though never 2); 
as a consequence the effective vertex
$Y_\chi \overline{\tilde{\ell}_L} \chi \ell_L$, 
is finite and generated radiatively~\footnote{Note that in the theory 
containing an extra
real scalar $\sigma$, one could also reason
differently to justify this result. Indeed, we could assign $\sigma$
LN equal to 0; then the
quartic term $\sigma \phi^\dagger \chi \tilde{\phi}$ fixes the $\chi$
LN also equal to 0, while the trilinear term
$\kappa {\rm Tr}\{(\chi^\dagger)^2\}$ breaks LN softly. 
In any case the discrete $Z_2$ symmetry guarantees that
the neutrino masses stay finite.}.
Despite the parallels of this discussion with models implementing a 
type II see-saw mechanism (which contain a tree-level $
\overline{\tilde{\ell}_L} \chi \ell_L$ coupling) there is an important
difference, namely, that in such theories both the expression
for the neutrino masses and the $0\nu\beta\beta$ decay amplitude
are {\em linear} in $\vevof\chi$; in contrast the corresponding results in 
our model are proportional to $ \vevof\chi^2 $ (see Eqs.~(\ref{eq:epsilon3})
and (\ref{eq:nu-mass-matrix}) below). 

The extension of the SM model we consider is related to the one presented in 
Refs.~\cite{Chen:2006vn,Chen:2007dc} as far as particle content
is concerned (we differ by adding the singlet $ \sigma $). However, the
symmetries and
quantum number assignments are different, which proves a crucial difference.
Had we included the hard term $\kappa \phi^\dagger \chi^\dagger \tilde{\phi}$
as was done in the above references, $\chi$ should have been assigned 
LN~$=2$; this would necessitate also including the tree-level
coupling $\overline{\tilde{\ell}_L} \chi \ell_L$ that would lead to the
usual type II see-saw scenario. In particular it would
have been inconsistent to assign LN zero to $ \chi$
or to arbitrarily exclude this Yukawa coupling from the Lagrangian,
for its coefficient would receive divergent radiative corrections;
in consequence the neutrino masses are not calculable.
If one requires $\chi$ to have LN equal to $0$,
as was done in these publications, the
quartic coupling $\kappa \phi^\dagger \chi^\dagger \tilde{\phi}$
must be absent; but then LN remains unbroken and the light
neutrino masses must vanish to all orders
which is again inconsistent with the results presented there. 

\subsection{The scalar spectrum\label{sec:scalar-spectrum}}

The requirement on the scalar potential of being bounded from below 
is fulfilled restricting the quartic couplings in Eq.~\eqref{eq:model.L} 
adequately; these conditions include~\footnote{Notice that the term 
$|\phi|^2\sigma^2$ is always positive.}
\begin{equation}
\lambda_{\sigma},\lambda_{\phi}>0 \,, \qquad
\lambda_{\phi\sigma} > - 2 \sqrt{\lambda_{\sigma}\lambda_{\phi}}~.
\label{eq:pot.stab}
\end{equation}
We have also checked that there is a non-trivial minimum 
on which the scalar neutral components acquire non-zero expectation values: 
$v_{\phi} \equiv \text{\ensuremath{\vevof{\phi^0}}} > 0$, 
$v_{\chi} \equiv \text{\ensuremath{\vevof{\chi^0}}} > 0$,
$v_{\sigma} \equiv \text{\ensuremath{\vevof{\sigma}}} > 0$, 
with  
\begin{equation}
\phi^{0}=\vevof{\phi^0}+\inv{\sqrt{2}}(\phi_{R}+i\phi_{I})\,,\qquad\chi^{0}=
\vevof{\chi^0}+\inv{\sqrt{2}}(\chi_{R}+i\chi_{I})\,,\qquad\sigma=
\vevof{\sigma}+\sigma_R \ .
\label{eq:field-shifts} 
\end{equation} 
In Appendix \ref{TripletVEV} we comment on the experimental limit 
on the scalar triplet VEV $v_{\chi}$, which we will find to be 
of the order of few GeV; to be conservative we will assume 
$v_{\chi}<5\,\mathrm{GeV}$, this satisfies 
$v_{\chi}\ll v_{\phi} \approx 174\, \textrm{GeV}$ where
$1/(v_{\phi}^{2}+2v_{\chi}^{2}) = g^{2}/2m_{W}^{2} = 2\sqrt{2}G_{F}$  
(derived from
Eq. \eqref{rho0} and the experimental limit on the $\rho$ parameter, 
$| \rho_0 - 1| \ll 1$~\cite{Nakamura:2010zzi}). 
In this approximation the minimization conditions can be easily solved: 
\begin{equation}
v_{\chi} \approx \frac{-\lambda_{6}v_{\sigma}v_{\phi}^{2}}{m_{\chi}^{2}+
v_{\phi}^{2}\lambda_{\phi\chi}+v_{\sigma}^{2}\lambda_{\sigma\chi}} \ , 
\qquad 
v_{\phi}^{2}\approx\frac{2\lambda_{\sigma}m_{\phi}^{2}-
\lambda_{\phi\sigma}m_{\sigma}^{2}}
{4\lambda_{\sigma}\lambda_{\phi}-\lambda_{\phi\sigma}^{2}} \ , 
\qquad 
v_{\sigma}^{2}\approx\frac{2\lambda_{\phi}m_{\sigma}^{2}-
\lambda_{\phi\sigma}m_{\phi}^{2}}
{4\lambda_{\sigma}\lambda_{\phi}-\lambda_{\phi\sigma}^{2}} \ .
\label{eq:triplet-vev}
\end{equation} 
Notice that the phase choice $\lambda_{6}<0$ 
is consistent  with $v_\chi$ being real and positive.
We set the $\sigma$ and $\phi$ mass squared terms 
in the Higgs potential negative to favour the development of such a 
minimum. Though we choose the $\chi$ mass squared term positive, this field
also acquires a small VEV induced by the 
doublet and singlet VEVs, similar to the case observed
in see-saw of type II models 
\cite{Konetschny:1977bn,Cheng:1980qt,Schechter:1980gr}. 

The scalar masses can be obtained by substituting 
Eq.~\eqref{eq:field-shifts} in the potential. 
Using the exact minimization conditions
to eliminate $m_{\phi}$ and $m_{\sigma}$ in favour of the VEVs, 
the mass terms for the charged scalars can be written  
\begin{eqnarray}
\lcal_{{\rm M}} & = & -\left(\kappa^{--}\;\; 
\chi^{--}\right)M_{D}^{2}\begin{pmatrix}\kappa^{++}\\
\chi^{++}
\end{pmatrix}-\left(\phi^{-}\;\; \chi^{-}\right)M_{S}^{2}\begin{pmatrix}\phi^{+}\\
\chi^{+}
\end{pmatrix} \ , \; {\rm with} \\
M_{D}^{2} & = & \begin{pmatrix}m_{\kappa}^{2}+
v_{\phi}^{2}\lambda_{\text{\ensuremath{\phi}}\kappa}+
v_{\sigma}^{2}\lambda_{\text{\ensuremath{\sigma}}\kappa}+
v_{\chi}^{2}\lambda_{\text{\ensuremath{\kappa\chi}}} & 2\mu_{\kappa}v_{\chi}\\
2\mu_{\kappa}v_{\chi} & 
m_{\chi}^{2}+
v_{\phi}^{2}(\lambda_{\phi\chi} + \lambda_{\phi\chi}^{\prime})
+v_{\sigma}^{2}\lambda_{\sigma\chi}+2v_{\chi}^{2}\lambda_{\chi}^{\prime}
 \end{pmatrix} \ , \\
M_{S}^{2} & = & \frac{\left(v_{\chi}\lambda_{\phi\chi}^{\prime}-
2v_{\sigma}\lambda_{6}\right)}
{2v_{\chi}}\left(\begin{array}{cc}
2v_{\chi}^{2} & -\sqrt{2}v_{\phi}v_{\chi}\\
-\sqrt{2}v_{\phi}v_{\chi} & v_{\phi}^{2} \end{array}\right) \ .
\end{eqnarray}
Analogously for the neutral sector 
\begin{eqnarray}
\lcal_{{\rm M}} & = & - \frac{1}{2} \left(\phi_{I} \;\; 
\chi_{I}\right)M_{I}^{2}\begin{pmatrix}\phi_{I}\\
\chi_{I}
\end{pmatrix}- 
\frac{1}{2} \left(\phi_{R} \;\; \chi_{R} \;\; 
\sigma_R \right)M_{R}^{2}\begin{pmatrix}\phi_{R}\\
\chi_{R}\\
\sigma_R 
\end{pmatrix} \ , \; {\rm with} \\
M_{I}^{2} & = & -\frac{v_{\sigma}\lambda_{6}}{v_{\chi}}\left(\begin{array}{cc}
4v_{\chi}^{2} & -2v_{\phi}v_{\text{\ensuremath{\chi}}}\\
-2v_{\phi}v_{\text{\ensuremath{\chi}}} & v_{\phi}^{2}
\end{array}\right) \ , \\
M_{R}^{2} & = & \left(\begin{array}{ccc}
4v_{\phi}^{2}\lambda_{\phi} & 2v_{\phi}\left(v_{\sigma}\lambda_{6}+
v_{\chi}\lambda_{\phi\chi}\right) & 2\sqrt{2}v_{\phi}
\left(v_{\sigma}\lambda_{\phi\sigma}+v_{\chi}\lambda_{6}\right)\\
2v_{\phi}\left(v_{\sigma}\lambda_{6}+v_{\chi}\lambda_{\phi\chi}\right) & 
4v_{\chi}^{2}(\lambda_{\chi}+\lambda^\prime_{\chi})-
v_{\phi}^{2}v_{\sigma}\lambda_{6}/v_{\chi} & 
\sqrt{2}\left(v_{\phi}^{2}\lambda_{6}+2v_{\sigma}v_{\chi}\lambda_{\sigma\chi}\right)\\
2\sqrt{2}v_{\phi}\left(v_{\sigma}\lambda_{\phi\sigma}+v_{\chi}\lambda_{6}\right) & 
\sqrt{2}\left(v_{\phi}^{2}\lambda_{6}+2v_{\sigma}v_{\chi}\lambda_{\sigma\chi}\right) & 
8v_{\sigma}^{2}\lambda_{\sigma}-{2v_{\phi}^{2}v_{\chi}\lambda_{6}}/{v_{\sigma}}
\end{array}\right) .
\end{eqnarray}
All eigenvalues of these mass matrices must be positive (except for the would-be 
Goldstone bosons providing the longitudinal vector boson degrees of freedom) 
in order to guarantee that the solution to the minimization conditions corresponds 
to a local minimum. 
This sets further constraints on the model parameters, which can be satisfied rather 
easily, especially in the limit $m_{\chi}\gg v_{\phi,\sigma,\chi}$.

Thus, we are left with two massive doubly-charged scalars $\kappa_{1,2}$, 
\begin{eqnarray}
\kappa_{1} & = & \text{\ensuremath{\cos\theta}}_{D} \kappa^{++}+
\sin\theta_{D} \chi^{++} \ , \nonumber \\
\kappa_{2} & = & - \text{\ensuremath{\sin\theta}}_{D} \kappa^{++}+
\text{\ensuremath{\cos\theta}}_{D} \chi^{++} \ , 
\label{eq:mixing-doubly}
\end{eqnarray}
with 
\begin{equation}
\sin2\theta_{D}=2\sin\theta_{D}\cos\theta_{D}=
\frac{4\mu_{\kappa}v_{\chi}}{m_{\kappa_{1}}^{2}-m_{\kappa_{2}}^{2}} \ , 
\label{eq:thetad}
\end{equation}
and only one massive, mainly triplet, singly-charged scalar $\omega$, 
\begin{equation}
\begin{array}{ccc}
\omega^{+} & = & - \text{\ensuremath{\sin\theta}}_{S}\phi^{+}+\cos\theta_{S}\chi^{+}
\end{array}\ , \; {\rm with} \quad \tan\theta_{S}=\frac{\sqrt{2}v_{\chi}}{v_{\phi}} \ .
\label{eq:mixing-singly}
\end{equation}
Similarly, there is a neutral scalar $A$ with imaginary components,  
\begin{equation}
\begin{array}{ccc}
A & = & - \text{\ensuremath{\sin\theta}}_{I}\phi_{I}+\cos\theta_{I}\chi_{I}
\end{array}\ , \; {\rm with} \quad \tan\theta_{I}=\frac{2v_{\chi}}{v_{\phi}} \ . 
\label{eq:mixing-neutral}
\end{equation}
There are also three neutral scalars along the real components
$\phi_R,~\chi_R,~\sigma_R$. We 
will denote these mass eigenfields by $h$ (mainly doublet), $H$ (mainly
triplet) and $s$ (mainly singlet). 
They are obtained rotating the current fields, what introduces other three 
mixing angles. 
Notice that in the limit $m_{\chi}\gg v_{\phi,\sigma}$, we have
$v_{\chi}\ll v_{\phi,\sigma}$ and 
$m_{{\kappa_{2}}, {\omega}, {A}} \approx m_{\chi}$, 
with all mixings small. 

\subsection{Some scalar couplings of phenomenological interest
\label{sec:RelevantCouplings}}

Once the quadratic terms of the Lagrangian are diagonalized we can read 
the interactions for the mass eigenfields. 
In the following we will need the scalar coupling to RH electrons 
\begin{equation}
\overline{e_{R a}^{c}}\ g_{ab}\ e_{R b} 
\left(\text{\ensuremath{\cos\theta}}_{D}\kappa_{1}-
\sin\theta_{D}\kappa_{2}\right)+\mbox{h.c.} \ , 
\label{eq:e-kappa-couplings}
\end{equation}
and the corresponding doubly-charged scalar couplings to gauge bosons
\begin{equation}
g^{2}\chi^{0\dagger}W_{\mu}^{-}W^{\mu-}\chi^{++}+\mbox{h.c.} \rightarrow 
g^{2}v_{\chi}W_{\mu}^{-}W^{\mu-}
\left(\text{\ensuremath{\sin\theta}}_{D}\kappa_{1}+
\cos\theta_{D}\kappa_{2}\right)+\mathrm{h.c.} \ ,
\label{eq:WW-kappaCouplings}
\end{equation}
as well as their trilinear couplings
\begin{equation}
-\mu_{\kappa}\kappa^{++}\left(\chi^{-}\right)^{2}
-2\mu_{\kappa}\kappa^{++}\chi^{--}\chi^{0\dagger}
+\lambda_{6}\sigma\left(\phi^{-}\right)^2\chi^{++}
+{\mathrm{h.c.}}\ ,
\label{eq:kappaScalarCouplings}
\end{equation}
which can be also expressed in terms of the mass eigenfields using 
Eqs.~(\ref{eq:mixing-doubly}--\ref{eq:mixing-singly}), and the corresponding
VEVs in Eq.~\eqref{eq:field-shifts}.
Finally, the Yukawa coupling changing charge and chirality writes 
\begin{equation}
\overline{\nu_{L}}\ Y_{e}\ e_R\,\phi^{+}+\mathrm{h.c.}=\ 
\overline{\nu_{L}}\ Y_{e}\ e_R\,\left(\text{\ensuremath{\cos\theta}}_{S} 
G^{+}-\sin\theta_{S} \omega^{+}\right)+\mathrm{h.c.} \ , 
\label{eq:SM-Yukawa-couplings}
\end{equation}
where $G^+= \cos\theta_{S}\phi^{+}+\sin\theta_{S}\chi^{+}$ is the would-be 
Goldstone boson providing the third component to the $W$.  

\section{Neutrinoless double beta decay \label{sec:0nu2beta}}

Both doubly-charged scalars $\kappa_{1, 2}$ have 
components along the singlet $\kappa^{++}$ and the triplet $\chi^{++}$. 
Therefore, they (respectively) contain couplings to RH electrons
and to $W$'s, generating an effective vertex $eeWW$ that mediates 
$0\nu\beta\beta$ decay. 
In this section we calculate this contribution to $0\nu\beta\beta$ 
decay and obtain the explicit constraints on the model parameters derived 
from the assumption that $0\nu\beta\beta$ decay will be observed in the next 
round of experiments. 

\begin{figure}
\begin{centering}
\includegraphics[width=0.55\columnwidth]{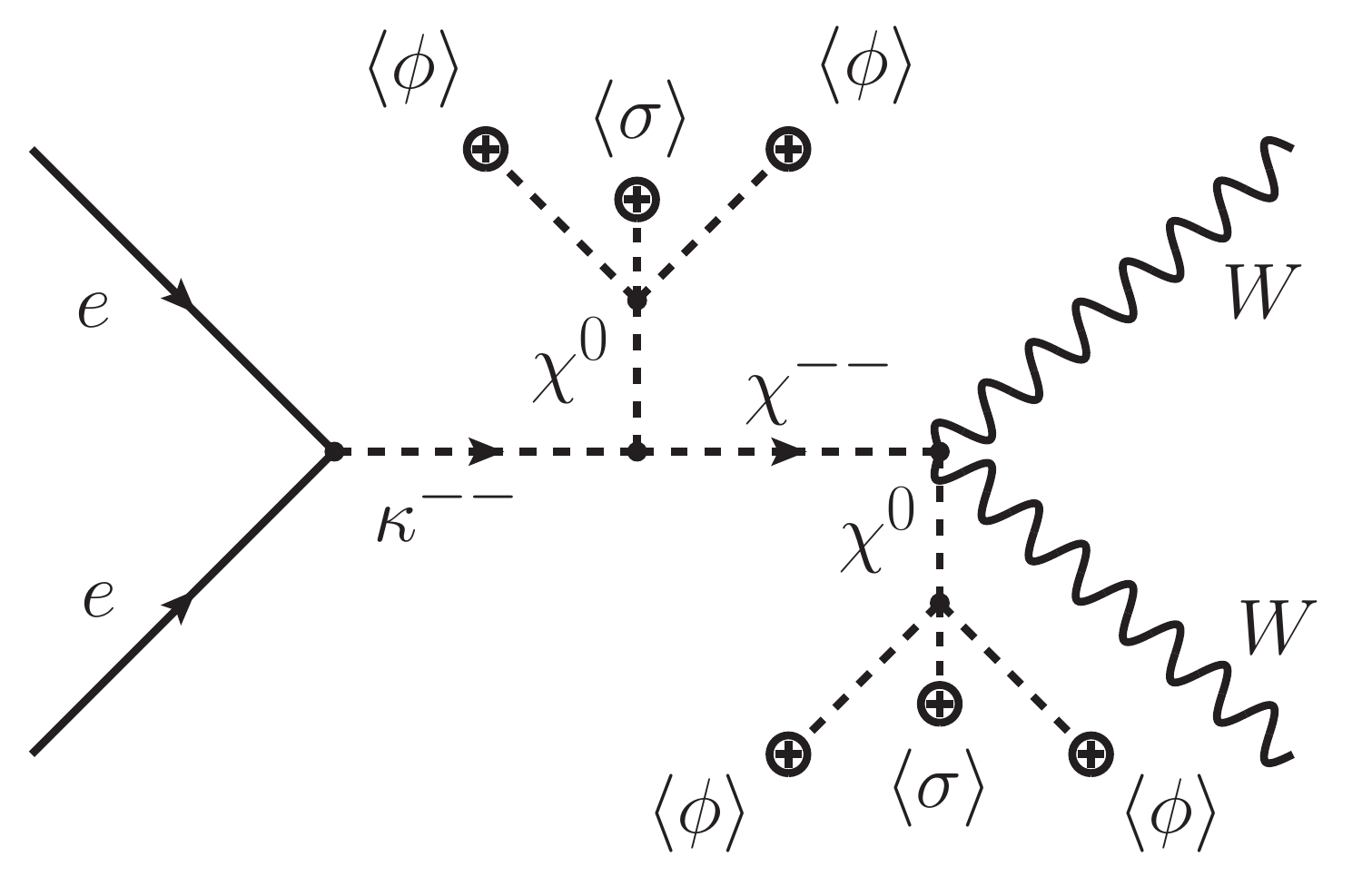}
\par\end{centering}
\caption{Dominant tree-level diagram contributing to the effective neutrinoless double
beta decay operator.
\label{fig:0nu2beta}}
\end{figure}

Assuming that $ m_{\kappa,\chi} \gg v_{\phi,\sigma } $ and
integrating out the heavy $\kappa$ and $\chi$ modes we find, after 
a straightforward calculation, that the effective Lagrangian contains the term
\begin{equation}
\mathcal{L}_{9}=
\frac{4(\lambda_6 v_\sigma)^2 \mu_{\kappa}  }{m_\kappa^2 m_\chi^6 }
\left(\overline{e_{R a}}\ g_{ab}^{*}\ e_{R b}^{c}\right)
\left(\phi^{\dagger}D^{\mu}\tilde{\phi}\right)
\left(\phi^{\dagger}D_{\mu}\tilde{\phi}\right)+\mathrm{h.c.} \ ,
\label{eq:derived.O9}
\end{equation}
as announced in the Introduction, and discussed in the companion paper 
\cite{delAguila:2011zz}. One can better understand  the origin
of this LNV interaction by considering the contribution of the dominant diagram
in Fig. \ref{fig:0nu2beta}, 
where the different couplings and VEVs involved are displayed explicitly. 
The corresponding $eeWW$ vertex at low energy ($q^{2}\ll m_{\kappa_{1,2}}^{2}$)
can be written as~\footnote{
It must be emphasized that the Lagrangian 
violating LN by 2 units for $0\nu\beta\beta $ decay
is proportional to $v_\chi^2$.
This must be compared with the linear dependence in $v_\chi$ obtained when 
the $\chi$ LN assignment is 2.
}
\begin{equation}
\mathcal{L}_{eeWW}=-\frac{2g^2\mu_{\kappa}v_{\chi}^{2}}
{m_{\kappa_{1}}^{2}m_{\kappa_{2}}^{2}} 
\overline{e_{R a}}\ g_{ab}^{*}\ e_{R b}^{c}\ W^{\mu}\ 
W_{\text{\ensuremath{\mu}}}+\mathrm{h.c.} \ ,
\label{eq:Langrangian-eeWW}
\end{equation}
where we have summed up all possible mass insertions in the internal 
propagator, and used Eq.~\eqref{eq:thetad}. 
This expression coincides with Eq.~\eqref{eq:derived.O9} when the scalar 
doublets develop a VEV in the
limit of large $m_{\kappa,\chi}$ ($\gg v_{\phi,\sigma}$), {\em i.e.},  
$v_\chi \approx -\lambda_6 v_\sigma v^2_\phi/m^2_\chi$ 
and $m_\kappa \approx m_{\kappa_1}$, $m_\chi \approx m_{\kappa_2}$. 

Let us particularize to the case $a=b=e$, relevant for $0\nu\beta\beta$,
and further integrate the two $W$'s to obtain the
appropriate 6-fermion contact interaction
\begin{equation}
\mathcal{L}_{0\nu\beta\beta}= 
\frac{G_{F}^{2}}{2m_{p}}\epsilon_3 \left(\bar{u}\gamma^{\mu}(1-\gamma_{5})d\right)
\left(\bar{u}\gamma_{\mu}(1-\gamma_{5})d\right)\bar{e}(1-\gamma_{5})e^{c}
\, ,
\label{eq:Lagrangian-0nu2beta}
\end{equation}
where $m_p$ denotes the proton mass and 
\begin{equation}
\epsilon_3=-\frac{8m_{p}\mu_{\kappa}v_{\chi}^{2}}{m_{\kappa_{1}}^{2}m_{\kappa_{2}}^{2}}g^*_{ee}\, .
\label{eq:epsilon3}
\end{equation}
This type of interactions has 
been already considered in the literature~\cite{Pas:2000vn}, where
limits from the most sensitive experiments at that moment were derived.
Since they have not been substantially improved, we will directly use
the results in~\cite{Pas:2000vn} from the Heidelberg-Moscow
experiment corresponding to $T_{1/2}>1.9\times 10^{25}\,\mathrm{years}$ which yields
\footnote{There is a misprint in Ref.~\cite{Pas:2000vn}. We thank the authors of this reference 
for providing us with the correct limit on $\epsilon_3$.} 
$\epsilon_3 < 1.4\times 10^{-8}$ at 90\%~C.L.
On the other hand, experiments  in the near future will be sensitive 
to lifetimes of the order of $6\times 10^{27}$ years~\cite{Barabash:2011fg}, {\em i.e.} 
a reduction factor on the coupling of roughly $0.05$. Then, in order to $0\nu\beta\beta$ 
decay be observable in the next 
round of experiments but still satisfy the present limits, we must require 
\begin{equation}
8.75\times10^{-11}\stackrel{\mathrm{{\scriptstyle Next}}}{<}
\frac{m_{p}\mu_{\kappa}v_{\chi}^{2}}
{m_{\kappa_{1}}^{2}m_{\kappa_{2}}^{2}}\; 
|g_{ee}|<1.75\times10^{-9}\quad(90\%\,\mathrm{C.L.})\,,
\label{eq:0nu2beta-bounds}
\end{equation}
where $m_{p}$ is the proton mass and the inequality with the 
superscript "$\mathrm{Next}$" 
corresponds to the requirement that $0\nu\beta\beta$ decay will be 
observed in the next generation of experiments \cite{Barabash:2011fg}. 
While the inequality without the 
superscript stands for the present experimental limit at the 90\%~C.L..
The conditions in Eq.~\eqref{eq:0nu2beta-bounds} will prove 
rather restrictive because its range of variation is relatively narrow. 
In fact, reducing the lower limit will appreciable enlarge the allowed 
parameter region as discussed below.

\begin{figure}
\begin{centering}
\includegraphics[width=0.49\columnwidth]{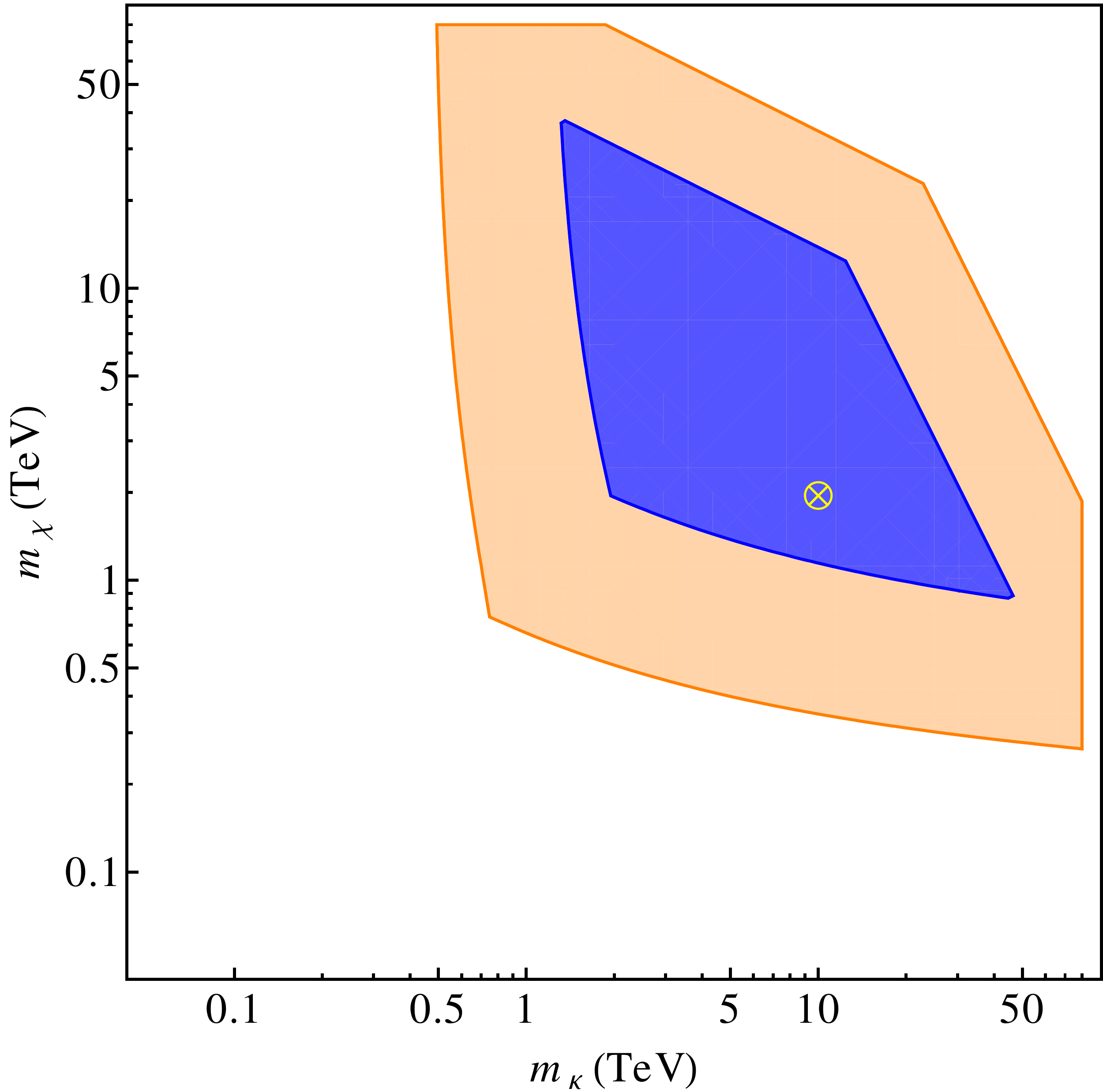}
\includegraphics[width=0.49\columnwidth]{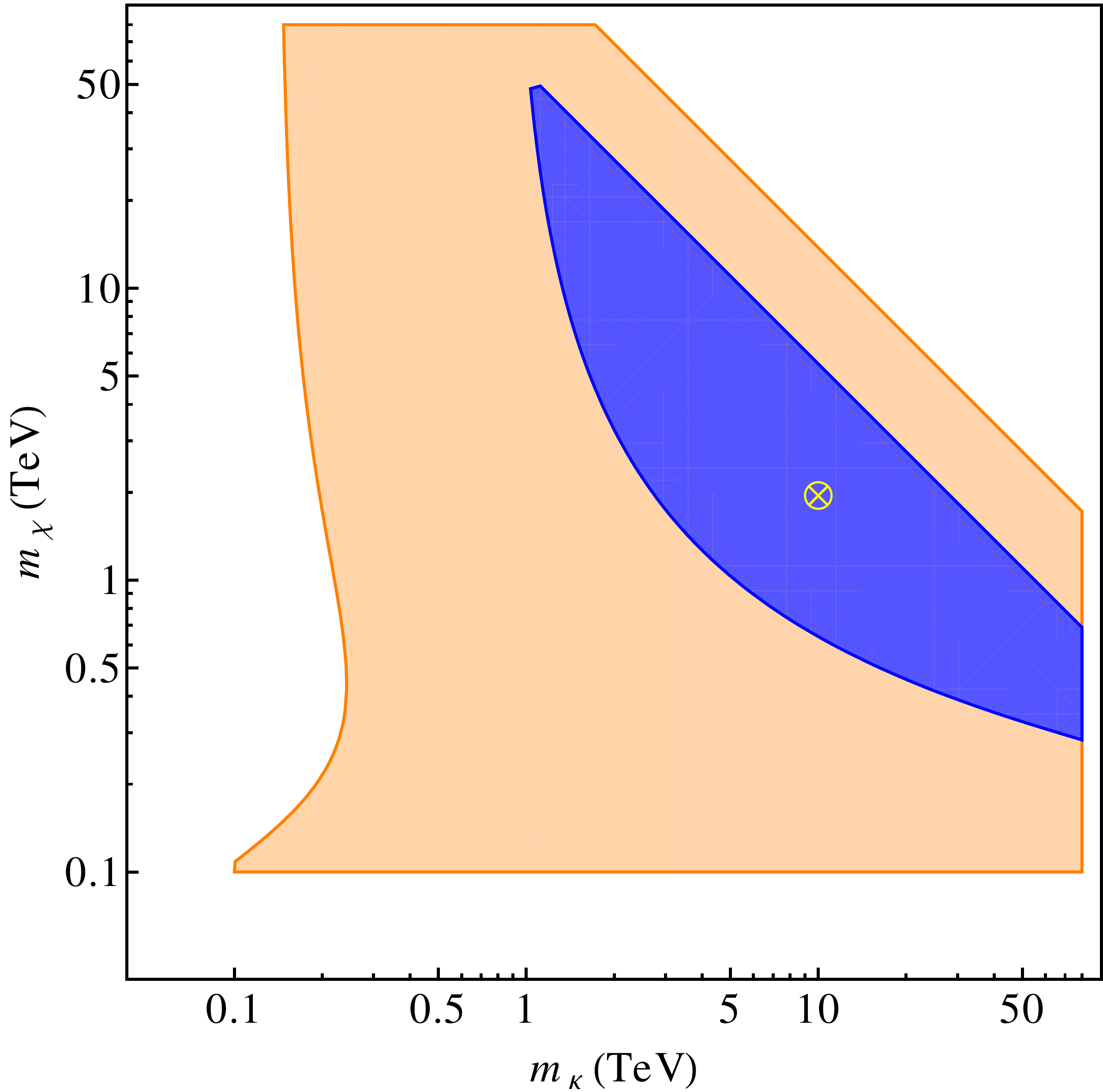}
\par\end{centering}
\caption{Projection on the $m_{\kappa}-m_{\chi}$ plane of the allowed 
parameter space region, with $m_{\kappa} \approx m_{\kappa_1}$ and 
$m_{\chi} \approx m_{\kappa_2}$. 
On the left (right) we draw the region assuming perturbative unitarity 
(a LN breaking scale $\mu_\kappa < 20$ TeV). 
The blue, darker (orange, lighter) areas correspond 
to $v_{\chi} = 2\ (5)\ {\mathrm {GeV}}$. 
The crosses stand for the reference point
$m_{\kappa_1} = 10\ {\mathrm {TeV}}, m_{\kappa_2} = 2\ {\mathrm {TeV}}$
(and $v_{\chi} = 2\ {\mathrm {GeV}}$, $\mu_{\kappa} = 15\ {\mathrm {TeV}}$,
with $g_{ee} = 1$ and $g_{e\mu} = 0.001$). 
\label{fig:plot-mk-mx}}
\end{figure}

Thus, the above lower limit together with the requirement of 
perturbative unitarity (indicated by the "Pert" superscript 
in the inequalities below) or naturality, which bounds from above the 
product of couplings and VEVs $\mu_{\kappa}v_{\chi}^{2}|g_{ee}|$,  
translate into an upper limit on the product of the scalar masses 
$m_{\kappa_{1,2}}$. 
These, however, are not precisely established 
because the perturbative bounds are in fact estimates 
that vary with the approach. 
In Fig. \ref{fig:plot-mk-mx} we show the allowed 
$m_{\kappa}-m_{\chi}$ region, where  
$m_{\kappa} \approx m_{\kappa_1}$ and 
$m_{\chi} \approx m_{\kappa_2}$ 
in the limit of small mixing angle $\theta_D$, 
for $v_{\chi} = 2$ (blue, darker) and 5 (orange, lighter) GeV
(see Appendix \ref{TripletVEV}), respectively, 
assuming perturbative unitarity (left) and a maximum 
LN breaking scale $\mu_\kappa$ (right). 
In the first case 
(see Eqs.~\eqref{eq:gs-ulimit} and \eqref{eq:muk-ulimit} in Appendix 
\ref{Perturbativityconstraints})
\begin{equation}
|g_{ee}| \stackrel{\mathrm{{\scriptstyle Pert}}}{<} \sqrt{4\pi}\, ,
\quad \mu_\kappa \stackrel{\mathrm{{\scriptstyle Pert}}}{<} 4\pi\ {\rm min}(m_{\kappa_{1,2}})\, ; 
\label{eq:pulimits}
\end{equation}
whereas in the second one 
\begin{equation}
|g_{ee}| \stackrel{\mathrm{{\scriptstyle Pert}}}{<} \sqrt{4\pi}\, ,\quad 
\mu_\kappa <  20\ {\rm TeV}\, . 
\label{eq:lnblimits}
\end{equation}
It must be noticed that all (pseudo-)observables violating LN 
are proportional to $\mu_{\kappa}v_{\chi}^{2}$. 
Hence, an increase in $v_{\chi}$ can be traded by the corresponding 
increase in $\mu_{\kappa}$, and vice-versa. 
So, the orange areas in Fig.~\ref{fig:plot-mk-mx} can be also 
interpreted as the allowed regions for $v_{\chi} = 2$ GeV and 
$\mu_\kappa \stackrel{\mathrm{{\scriptstyle Pert}}}{<} 25\pi\ {\rm min}(m_{\kappa_{1,2}})$ 
(left) and $\mu_\kappa < 125$ TeV (right). 
One may wonder at this point why we choose
the bound of 20 TeV for $\mu_\kappa$ in Eq.~\eqref{eq:lnblimits};
or equivalently, what is the effect of varying
such a value. The answer is simple. The blue,~darker
region in the right panel of Fig.~\ref{fig:plot-mk-mx} disappears for
$\mu_\kappa \lesssim 8$ TeV, which only reflects the narrowness
of the range allowed by Eq.~\eqref{eq:0nu2beta-bounds}, as required
by our main working assumption that $0\nu\beta\beta$
decay will be observed in the next round of experiments.
The allowed regions in Fig.~\ref{fig:plot-mk-mx} are appreciably enlarged
by reducing the lower limit in this equation.
This can also be achieved by further
increasing $\mu_\kappa$.

These areas are also bounded from below due to the non-observation 
of doubly-charged scalars; we can then assume $m_{\kappa_{1,2}} > 100$ GeV, 
as discussed in Section \ref{sec:colliders}. 
However, the enclosed areas in Fig. \ref{fig:plot-mk-mx} are 
further reduced by a more stringent and subtle constraint. 
As we shall discuss below, bounds on LFV processes 
(see section \ref{sec:Lepton-flavour-violation}) like 
$\tau^{-}\rightarrow e^{+}\mu^{-}\mu^{-}$ banish $m_\kappa$ 
to large values if the corresponding coupling product 
$g_{\tau e}g_{\mu \mu}^*$ is sizeable, which is required
because neutrino masses are 
proportional to $g_{ab}$
(see section \ref{sec:the neutrino mass}), 
and $g_{\tau e}$ and $g_{\mu \mu}$ 
must be large in order to accommodate the observed neutrino spectrum. 
Moreover, both $m_\kappa$ and $m_\chi$ enter in the two-loop 
integrals generating neutrino masses, but these tend to zero 
in the limit $m_\chi / m_\kappa \rightarrow 0$. 
As a result, both scalar 
masses are constrained, but differently, by the 
$\tau^{-}\rightarrow e^{+}\mu^{-}\mu^{-}$ bound. 
The regions in
Fig. \ref{fig:plot-mk-mx} satisfy all experimental restrictions, 
including the upper bound in Eq. (\ref{eq:0nu2beta-bounds}). 
The LHC will further reduce the allowed regions, 
mainly in the case of large LN breaking scale $\mu_\kappa$.  
We also provide a ``benchmark point'', denoted by a cross
in the figures, where all constraints are satisfied,
and which we will use as reference throughout the paper.
As can be deduced from this Figure the parameter space 
is rather constrained in this simple model when we require that 
the values of couplings and scalar masses stay natural, but one can 
think of other models within this class 
of theories where these constraints are significantly relaxed
(at the price of complicating the spectrum through the introduction 
of additional scalars).

\section{Lepton flavor violation constraints\label{sec:Lepton-flavour-violation} }

We will show in the next section that
in order to obtain neutrino masses in agreement with experiment, the
doubly-charged scalar Yukawa couplings $ g_{ab} $ and the
$m_{\chi} / m_{\kappa}$ ratio cannot be too small. 
In such a case some of the predicted LFV rates 
can become large enough to be at the verge of their 
present experimental bounds, especially for very rare processes like 
$\mu^{-}\rightarrow e^{+}e^{-}e^{-}$
or $\mu^{-}\rightarrow e^{-}\gamma$. Thus, we can use LFV 
processes to further constrain the model, and perhaps to confirm 
or exclude it in the near future. 

In this section we will briefly discuss the most 
restrictive process $\mu^{-}\rightarrow e^{+}e^{-}e^{-}$, 
whose tree-level amplitude is obtained by the 
single exchange of the doubly-charged 
scalar $\kappa$. 
The corresponding branching ratio equals 
\begin{equation}
\mathrm{BR}(\ell_{a}^{-}\rightarrow\ell_{b}^{+}\ell_{c}^{-}\ell_{d}^{-})=
\frac{1}{2(1+\delta_{cd})}
\abs{\frac{g_{ab}g_{cd}^{*}}{G_{F}\tilde{m}_{\kappa}^{2}}}^{2}
\mathrm{BR}(\ell_{a}^{-}\rightarrow\ell_{b}^{-}\nu\bar{\nu})\,, 
\label{eq:Rmuto3e}
\end{equation}
where $\delta_{cd}$ takes into account the fact that there may be 
two identical particles in the final state, as in our case, 
and 
\begin{equation}
\frac{1}{\tilde{m}_{\kappa}^{2}}\equiv\frac{\cos^{2}\theta_{D}}
{m_{\kappa_{1}}^{2}}+\frac{\sin^{2}\theta_{D}}{m_{\kappa_{2}}^{2}}\,.
\label{eq:eff-mkmass}
\end{equation}
(If $\sin^{2}\theta_{D}\ll1$, 
the effective mass $\tilde{m}_{\text{\ensuremath{\kappa}}}\approx m_{\kappa_{1}}$ 
since we, in practice, assume that
$m_{\kappa_{1,2}}$ are never very different.) 
Then, the current experimental limit on 
$\mathrm{BR}(\mu^{-}\rightarrow e^{+}e^{-}e^{-}) < 1.0\times10^{-12}$ 
\cite{Nakamura:2010zzi} translates into 
\begin{equation}
|g_{\mu e}g_{ee}^{*}|<2.3\times10^{-5}\,(\tilde{m}_{\text{\ensuremath{\kappa}}}/\mathrm{TeV})^{2}, 
\label{mnuemu}
\end{equation}
which is mainly a constraint on $g_{\mu e}$ because $g_{ee}$ must be  
relatively large if $0\nu\beta\beta$ decay has to be observable at the
next generation of experiments. 

Related processes provide weaker constraints. Thus, 
$\mu^{-}\rightarrow e^{-}\gamma$ proceeds at one loop and is 
suppressed 
by the corresponding loop factor, 
and similarly for $\mu-e$ conversion in nuclei. 
The bounds from $\mu^{+}e^{-}\longleftrightarrow\mu^{-}e^{+}$ 
(muonium-antimuonium conversion) or muon-positron conversion, 
although tree-level processes, are also 
less restrictive (for a discussion of LFV processes mediated by doubly-charged 
scalar singlets in a similar model see \cite{Nebot:2007bc,Raidal:1997hq}).  
All these processes and the analogous ones involving 
$\tau$ leptons, as well as the 
corresponding (anomalous) magnetic moments will 
be discussed in detail with more generality elsewhere. 

Here we shall be mainly interested in the interplay of 
a large $0\nu\beta\beta$ decay rate and a realistic pattern 
of Majorana masses, and for this purpose it is sufficient to show the restrictions 
on these (pseudo-)observables in simple SM extensions as 
the one at hand, 
and indicate which further processes may be within the reach 
of new experiments.
In our case, the most restrictive process besides 
$\mu^{-}\rightarrow e^{+}e^{-}e^{-}$ is $\tau^{-}\rightarrow e^{+}\mu^{-}\mu^{-}$, 
which will be discussed below taking into account the neutrino mass 
requirements.

\section{Neutrino mass generation and $\theta_{13}$ expectation 
\label{sec:the neutrino mass}}

In the model under consideration LN is not conserved
when the couplings $\mu_{\kappa}$, $\lambda_{6}$ , $g_{ab}$
and $Y_{e}$ are non-vanishing. In this case there is
no protection against the neutrinos acquiring Majorana masses $m_\nu$,
which will then be proportional to all four couplings,
are finite, and 
appear at the two-loop level, as explained in Section \ref{sec:The-model} 
and shown by explicit calculation in Appendix \ref{sec:loop-integrals}.
(If any of these couplings
vanishes a conserved 
lepton number remains after spontaneous symmetry breaking
and the neutrino masses will vanish.) These masses are generated
by the non-renormalizable interaction 
$\sigma \overline{\tilde{\ell}_L} \chi \ell_L$ generated at two loops, 
when $\sigma$ and  $\chi$ develop VEVs;
the corresponding coupling being $m_\nu / v_\sigma v_\chi$.
In contrast, the see-saw type II
coupling $\overline{\tilde{\ell}_L} \chi \ell_L$ violates
the $Z_2$ symmetry and is forbidden to all orders.

\begin{figure}
\begin{centering}
\includegraphics[width=0.5\columnwidth]{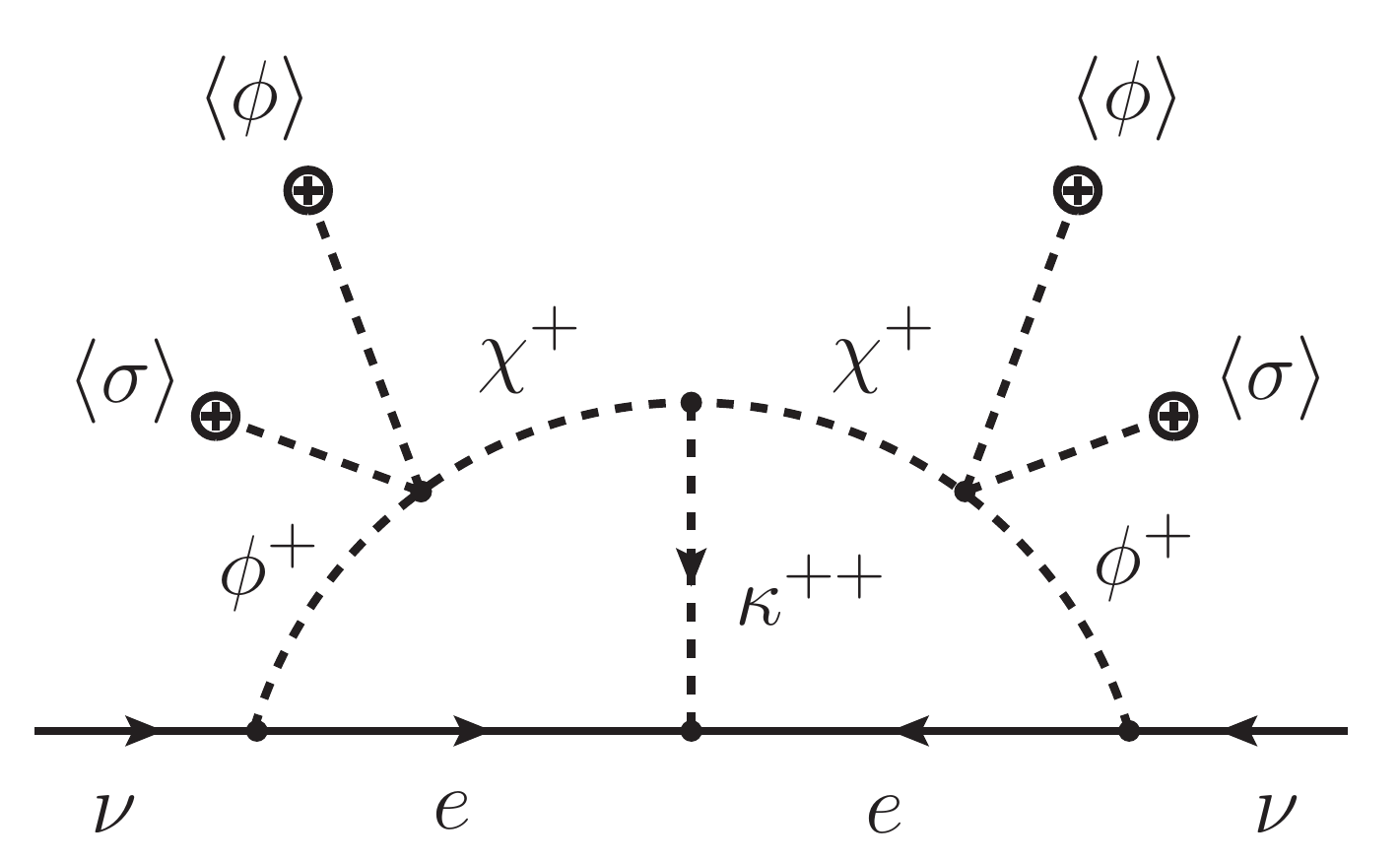}
\par\end{centering}
\caption{Two-loop diagram contributing to neutrino masses. 
\label{fig:numass-insertions}}
\end{figure}

In Fig.~\ref{fig:numass-insertions} we draw one of the diagrams. 
Defining the neutrino mass matrix as usual, 
\begin{equation}
\mathcal{L}_{m}=-\frac{1}{2}\overline{\nu_{L}}m_{\nu}\nu^c_{L}+\mathrm{h.c.} \ ,
\label{eq:nu-mass-lagrangian}
\end{equation}
we can write, taking $v_{\chi}\ll v_{\phi}$ (see Eq.~\eqref{eq:bound-triplet-vev} and 
Appendix \ref{TripletVEV}),
\begin{equation}
\left(m_{\nu}\right)_{ab}=\frac{\mu_{\kappa}v_{\chi}^{2}}{2(2\pi)^{4}
v_{\phi}^{4}}m_{a}g^*_{ab}m_{b}I_{\nu} \ ,
\label{eq:nu-mass-matrix}
\end{equation}
where $I_{\nu}$ is the sum of the (rescaled) loop
integrals from the different graphs~\footnote{
Again we note that neutrino masses, which violate LN by 2 units, are
proportional to $v_\chi^2$ in our case.
In contrast the dependence is linear in the estimate for the model in 
Refs.~\cite{Chen:2006vn,Chen:2007dc}, showing that $\chi$ must be assigned LN  2, 
as in see-saw models of type II.
Besides, the neutrino masses are in fact infinite in that particular case, a point obscured because divergent diagrams were omitted.
}. $I_{\nu}$ can be estimated in the mass 
insertion approximation with $m_{\kappa,\chi} \gg m_W$. For instance, 
in this limit one of the contributions, $I_{1}$, corresponding to the diagram in 
Fig. \ref{fig:numass-insertions} gives (neglecting the lepton masses in the denominator 
and assuming equal masses for the doubly and singly-charged triplet components)
\begin{equation}
I_{1}=(4\pi)^{4}m_{\chi}^{4}\int \frac{k\cdot q}
{k^{4}(k^{2}-m_{\chi}^{2})q^{4}(q^{2}-m_{\chi}^{2})((k-q)^{2}-m_{\kappa}^{2})} \ .
\end{equation}
In this approximation the full $I_{\nu}$ is a dimensionless 
function of $y=(m_{\chi}/m_{\kappa})^{2}$ 
of order one, except for $y\rightarrow0$, in which case it tends to zero 
as $y\log y$. 
In contrast with $I_1$ which tends faster to zero,
as $(y\log y)^2$, for vanishing $y$. 
$I_{\nu}$ is also bounded from above, going to a constant 
of order $2$ for $y \rightarrow \infty$.  
A complete calculation taking into 
account the $W$-mass, 
as well as the $v_{\chi}$ corrections and the new scalar
mass scales, is presented in Appendix~\ref{sec:loop-integrals}. 
A reasonable approximation is to neglect higher $v_{\text{\ensuremath{\chi}}}$
effects and take all triplet masses equal 
$m_{\kappa_{2}}=m_{\omega}=m_{A}$; in fact, in the physical limit 
$v_{\chi}\ll v_{\phi}$, $\kappa_{2}$ is mainly 
the doubly-charged triplet component, $\omega$ the singly-charged 
one, and $A$ the imaginary part of the neutral triplet component.

In our model the Yukawa couplings $ g_{ab} $ appear 
in the neutrino mass matrix, the  $0\nu\beta\beta$ decay amplitude and
the amplitudes for the LFV processes, and this translates into 
rather stringent constraints 
on the allowed neutrino mass matrices  (once 
one insists in dealing with a perturbative theory 
up to several tens of TeV). We now consider these constraints.

Assuming $\mu_{\kappa}\sim10\,\mathrm{TeV}$ and 
$v_{\chi} < 2~\mathrm{GeV}$, and taking 
$I_{\nu}, |g_{ee}|\sim1$, Eq.~(\ref{eq:nu-mass-matrix}) 
gives $|\left(m_{\nu}\right)_{ee}|\sim 3.7\times10^{-6}\,\mathrm{eV}$. 
How large can it be in general ? 
Using perturbativity limits \eqref{eq:pulimits} and $I_{\nu}\sim1$, we get 
\begin{equation}
|\left(m_{\nu}\right)_{ee}|\stackrel{\mathrm{{\scriptstyle Pert}}}{<}
1.6\times10^{-5}\ \left(\frac{{\rm min}(m_{\kappa_{1,2}})}{\mathrm{TeV}}\right)\,
\mathrm{eV}\stackrel{\mathrm{{\scriptstyle Next,Pert}}}{<}1.6\times10^{-4}\,
\mathrm{eV},
\label{eq:mee-pert-bound}
\end{equation}
where the upper limit is obtained by taking 
${\rm min}(m_{\kappa_{1,2}}) \sim 10\,\mathrm{TeV}$ (see Fig. \ref{fig:plot-mk-mx}, left). 
Alternatively, we can translate the limits on $0\nu\beta\beta$ decay 
in Eq.~\eqref{eq:0nu2beta-bounds} into bounds on $\left(m_{\nu}\right)_{ee}$, 
but for large scalar masses this limit is less stringent than \eqref{eq:mee-pert-bound}. 
In either case, $|\left(m_{\nu}\right)_{ee}|$ is typically less than $10^{-4}$.

There is in addition a quite strong bound on $\left(m_{\nu}\right)_{e\text{\ensuremath{\mu}}}$
from $\mu\rightarrow eee$. Substituting Eq. \eqref{mnuemu} in the generic neutrino mass 
expression in Eq.~\eqref{eq:nu-mass-matrix}  we find
\begin{equation}
|\left(m_{\nu}\right)_{e\mu}|<2.3\times10^{-5}
\left(\frac{\tilde m_{\kappa}}{\mathrm{TeV}}\right)^{2}
\frac{\mu_{\kappa}v_{\chi}^{2}}{2(2\pi)^{4}v_{\phi}^{4}}
\frac{m_{e}m_{\mu}}{|g_{ee}|}I_{\nu} \ .
\end{equation}
The constraint that a signal is seen in near-future $ 0 \nu\beta\beta $ 
decay experiments (left inequality in Eq.~\eqref{eq:0nu2beta-bounds})
can then be used to eliminate $ |g_{ee}|$; in this way
we obtain 
\begin{equation}
|\left(m_{\nu}\right)_{e\mu}|\stackrel{\mathrm{{\scriptstyle Next}}}{<}
4.3 I_{\nu} \frac{\tilde m_{\kappa}^2 \mu_{\kappa}^{2}}{m_{\kappa_{1}}^{2}m_{\kappa_{2}}^{2}} 
\frac{v_{\chi}^{4}}{v_{\phi}^{4}}\,\mathrm{eV} 
\stackrel{\mathrm{{\scriptstyle Next,Pert}}}{<} 1.2\times10^{-5}\,\mathrm{eV} \ ,
\label{eq:mem.bound}
\end{equation}
where in the second inequality we used $I_{\nu} \sim 1$, $v_{\chi} = 2\,\mathrm{GeV}$, 
the naturality limit on $\mu_{\kappa}$ (Eq.~\eqref{eq:pulimits}), 
and $\sin\theta_{D} \ll 1$. 
If we had used the LN breaking scale $\mu_{\kappa} = 10$ TeV and the production 
limit on $m_{\kappa_2} > 0.1$ TeV, then $|\left(m_{\nu}\right)_{e\mu}| < 7\times10^{-4}$.

The final result of this phenomenological discussion is that both, $|\left(m_{\nu}\right)_{ee}|$
and $|\left(m_{\nu}\right)_{e\mu}|$, must be below $\sim10^{-4}\,\mathrm{eV}$,
and this follows from 
requiring that {\it (i)} $0\nu\beta\beta$ decay is at the reach of 
the next round of experiments, and {\it (ii)} that the theory is perturbative and free of 
unnatural fine tuning up to several tens of TeV. 
These limits could be somewhat relaxed: in the $|\left(m_{\nu}\right)_{ee}|$ case by making 
doubly-charged scalar masses larger, and in the $|\left(m_{\nu}\right)_{e\mu}|$ case by allowing for 
a smaller $|g_{ee}|$. However, this is at the price of generating some tension with the naturality 
constraint in the former case, and spoiling the possibility of observing $0\nu\beta\beta$ 
decay induced by scalars in the near future in the latter one. 
There are additional but less severe bounds on the remaining $g_{ab}$ from other 
LFV processes; we discuss 
them below, when presenting
the plots for the relevant neutrino mass (pseudo-)observables 
satisfying present experimental restrictions. 

\subsection{Prediction for the third neutrino mixing angle $\theta_{13}$
\label{sec:theta13}}

The question now becomes whether it is possible to accommodate the
observed spectrum of neutrino masses and mixing angles in this type of models 
once the above experimental constraints are imposed. 
In the following we will use the standard parameterization 
of the neutrino 
mass matrix~\cite{Bilenky:1980cx,Schechter:1980gr,Bilenky:1987ty,Nakamura:2010zzi} in terms
of 3 mass parameters, 3 mixing angles and 3 phases: 
\begin{equation}
m_{\nu}=UD_{\nu}U^{T}\; , \quad {\rm with}\;\; D_{\nu}=\mathrm{diag}(m_{1},m_{2},m_{3}) 
\label{masses}
\end{equation}
and
\begin{eqnarray}
U=\left(\begin{array}{ccc}
c_{13}c_{12} & c_{13}s_{12} & s_{13}e^{-i\delta}\\
-c_{23}s_{12}-s_{23}s_{13}c_{12}e^{i\delta} & c_{23}c_{12}-s_{23}s_{13}s_{12}e^{i\delta} & s_{23}c_{13}\\
s_{23}s_{12}-c_{23}s_{13}c_{12}e^{i\delta} & -s_{23}c_{12}-c_{23}s_{13}s_{12}e^{i\delta} & c_{23}c_{13}
\end{array}\right)\left(\begin{array}{ccc}
e^{i\alpha_{1}/2}\\
 & e^{i\alpha_{2}/2}\\
 &  & 1
\end{array}\right)\ ,
\label{UPMNS}
\end{eqnarray} 
where $s_{ij} \equiv \sin\theta_{ij}$ and $c_{ij} \equiv \cos\theta_{ij}$. 
A global fit to neutrino oscillation data gives (see, for instance, \cite{Schwetz:2011zk})
$\Delta m_{21}^{2} \equiv m_{2}^{2}-m_{1}^{2}=(7.59{+0.20\atop -0.18})\times10^{-5}\:\textrm{eV}^{2}$,
$\Delta m_{31}^{2} \equiv m_{3}^{2}-m_{1}^{2}=\:(2.50{+0.09\atop -0.16})\times10^{-3}\:\textrm{eV}^{2}$,
$s_{12}^{2}=0.312{+0.017\atop -0.015}$, $s_{23}^{2}=\,\,0.52{+0.06\atop -0.07}$, 
$s_{13}^{2}=\,0.013{+0.007\atop -0.005}$. 
Neutrino oscillations are not sensitive to the phases $\alpha_{1}$ and $\alpha_{2}$, 
nor to a common mass scale which is conventionally chosen to be the lightest 
neutrino mass. 
$\delta$, which appears multiplied by $s_{13}$, is beyond
present experimental sensitivity.  
The sign of $\Delta m_{31}^{2}$ is not presently known, and could be 
negative (known as inverted hierarchy), however, in this case
$|\left(m_{\nu}\right)_{ee}|>10^{-2}\,\mathrm{eV}$ and 
cannot be accommodated within our model; we 
will therefore consider only the normal hierarchy case 
$\Delta m_{31}^{2} > 0 $. 
Finally, recent data on electron neutrino appearance at T2K~\cite{Abe:2011sj} 
and Double Chooz  \cite{dc-3393-v3:2011} experiments point out to a
mixing angle $\theta_{13}$ different from zero.

A possible way of identifying the allowed region in parameter space would
be to first generate random values for masses, angles and phases within 
the 1~$\sigma$ regions experimentally allowed in Eqs.~\eqref{masses} 
and \eqref{UPMNS}, and obtain scatter plots for $m_\nu$. 
Then, using Eq.~\eqref{eq:nu-mass-matrix} we can solve for 
$g_{ab}$, up to an overall factor $ \propto \mu_\kappa v_\chi^2  I_\nu/v_\phi^4$,
and then find the values of $\mu_\kappa,v_\chi$, $m_\kappa,m_\chi$
which respect the constraints discussed in the previous sections. 
The potential problem we face is due to the 
specific form of the neutrino mass matrix, which contains $m_a g^*_{ab} m_b$ 
(see Eq.~\eqref{eq:nu-mass-matrix}), and is therefore suppressed
for the first generations due to the 
light charged-lepton mass factors. To compensate this may 
require $g_{ab}$ to be too large to meet the 
bounds required by $0\nu\beta\beta$ decay (Section~\ref{sec:0nu2beta}),  
LFV processes (Section~\ref{sec:Lepton-flavour-violation})
and perturbative unitarity (Appendix~\ref{Perturbativityconstraints}).
An alternative way to proceed is noticing that in practice
(see (\ref{eq:mee-pert-bound}) and (\ref{eq:mem.bound}))
we are asking if $|(m_\nu)_{ee, e\mu}| \sim 0$ is 
consistent with neutrino oscillation data (a question also of general interest 
not only within the model under consideration). 
These additional constraints will hold only within a limited region of the allowed
neutrino masses and mixing parameters, which then implies that the type of
models under consideration gives rather clear predictions about some of
these parameters. 

\begin{figure}
\begin{centering}
\includegraphics[width=0.55\columnwidth]{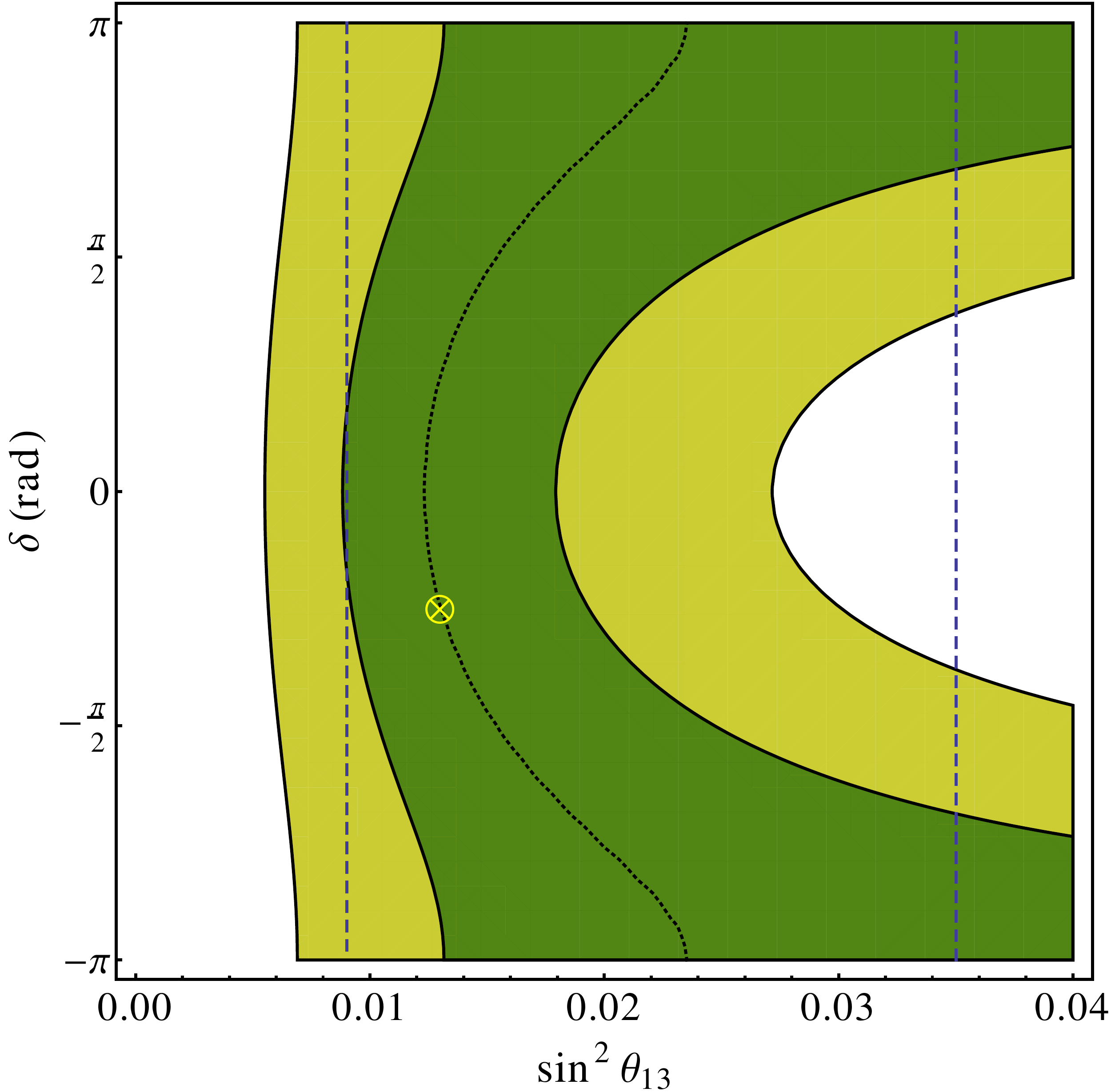}
\end{centering}
\caption{Allowed $\sin^2\theta_{13}-\delta$ region for  $|(m_{\nu})_{ee, e\mu}|=0$. 
The green,~darker region is obtained when measured mixings and mass differences 
(except $\sin\theta_{13}$) are varied within 1~$\sigma$; while the yellow,~lighter one 
is obtained by varying them within 3~$\sigma$. The middle dotted curve corresponds 
to the central values of the neutrino masses and mixings in the global fit performed in 
Ref.~\cite{Schwetz:2011zk}. For comparison, we also draw the recent Double Chooz 
\cite{dc-3393-v3:2011} 1~$\sigma$ limits, where statistical and systematic errors are 
added in quadrature (vertical dashed lines).
The cross stands for the reference point in Fig. \ref{fig:plot-mk-mx}.
\label{fig:graf-delta-s13}}
\end{figure}

In order to see how this comes about it is useful to go through
a straightforward parameter counting exercise:
$m_\nu $ is a $3\times 3$  complex and symmetric matrix specified
by 12 real numbers: 3 of these are unphysical and can be absorbed 
in re-phasing the neutrino fields, and 5 of the remaining 9 are measured
(2 mass differences and 3 mixing angles, where we include $\theta_{13}$).
If we now impose $(m_\nu)_{ee, e\mu} = 0$, corresponding
to 4 additional (real) constraints, only a set of
points (or narrow regions, allowing for experimental accuracy)
will be consistent. In fact, there may be no allowed values at all ! 
We have checked, however, that for each allowed choice of the experimentally
measured parameters there is a unique solution
for $ \alpha_1,~ \alpha_2,~ \delta $ and $m_1$ satisfying
all these restrictions. For example
using the central values of the global fit given after Eq.~(\ref{UPMNS}) we find:
\begin{equation}
\alpha_{1}=0.65\quad,\quad\alpha_{2}=-2.32\quad,
\quad\delta=-0.78\quad,\quad m_{1}=0.0036\,\mathrm{eV} \ , 
\end{equation}
with 
\begin{equation}
m_{\nu}\approx10^{-2}\left(\begin{array}{ccc}
 0 & 0 & 0.59 +0.58 i \\
 0 & 2.47 -0.2 i & 2.64 +0.2 i \\
 0.59 +0.58 i & 2.64 +0.2 i & 2.12 -0.21 i
\end{array}\right)\,\mathrm{eV} \ .
\label{eq:mnu-example}
\end{equation}
This exercise can be repeated for different values of $\sin^2\theta_{13}$, 
and the amazing result is that for the central values of 
$ \Delta m^2_{21, 31}$ and $s_{12, 23}^{2}$ 
there are solutions only for $0.012<\sin^2\theta_{13}<0.024$, a 
range of values that roughly coincides with the result 
obtained by the global fits to present data and by recent T2K and Double Chooz experiments. 
To illustrate this result we present in Fig. \ref{fig:graf-delta-s13} 
the $\sin^2\theta_{13}-\delta$ region allowed when $|(m_{\nu})_{ee, e\mu}|=0$ is imposed. 
The green,~darker region is obtained when measured mixings and mass differences 
(except $\sin\theta_{13}$) are varied within 1~$\sigma$, while the yellow,~lighter one 
is obtained 
by varying them within 3~$\sigma$. For comparison we also present the recent 
Double Chooz result~\cite{dc-3393-v3:2011} ($\sin^2(2\theta_{13})=0.085\pm 0.029\pm 0.042$, 
adding statistical and systematic 
errors quadratically we obtain $\sin^2\theta_{13}=0.022\pm 0.013$) as a vertical band, 
while the cross stands for the reference point 
in Fig.~\ref{fig:plot-mk-mx}.     

\begin{figure}
\begin{centering}
\includegraphics[width=0.7\columnwidth]{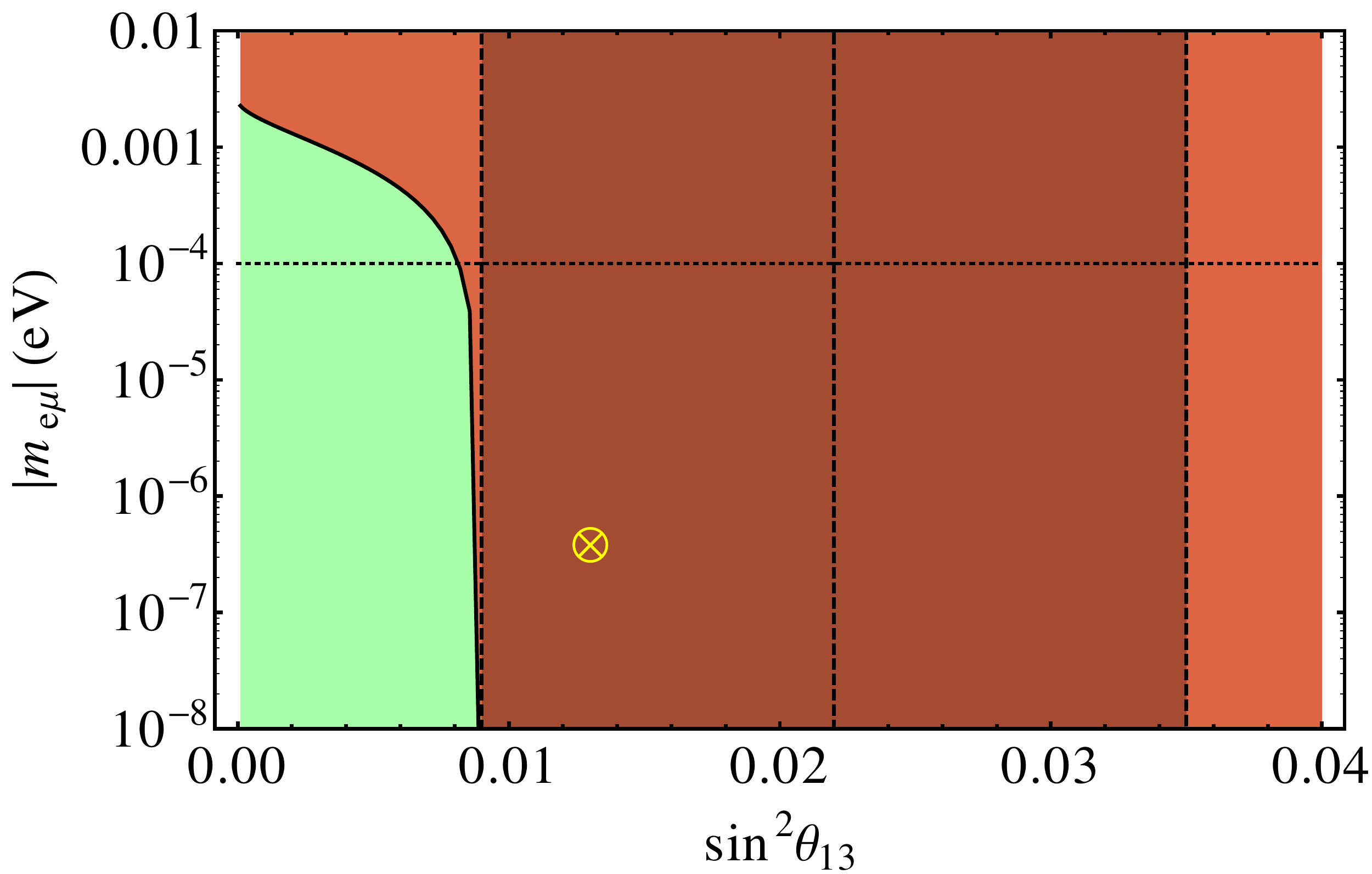}
\par\end{centering}
\caption{$|(m_{\nu})_{e\mu}|$ as a function of $\sin^2\theta_{13}$.
The green,~lighter area which extends to all the panel is obtained varying 
the other measured masses and mixings within 1~$\sigma$.  
The red, darker region satisfies the extra constraint 
$|(m_{\nu})_{ee}| < |(m_{\nu})_{e\mu}|$. 
The cross corresponds to the reference point in Fig. \ref{fig:plot-mk-mx}, and can be reached
with either condition  $|(m_{\nu})_{ee}| {<\atop >} |(m_{\nu})_{e\mu}|$;
whereas the darkest vertical 1~$\sigma$ band stands for the recent Double Chooz result, 
summing errors in quadrature. 
We also draw the line $|(m_{\nu})_{e\mu}| = 10^{-4}$, which is the upper-limit estimate 
in this model.
\label{fig:s13-mem}}
\end{figure}

\begin{figure}
\begin{centering}
\includegraphics[width=0.7\columnwidth]{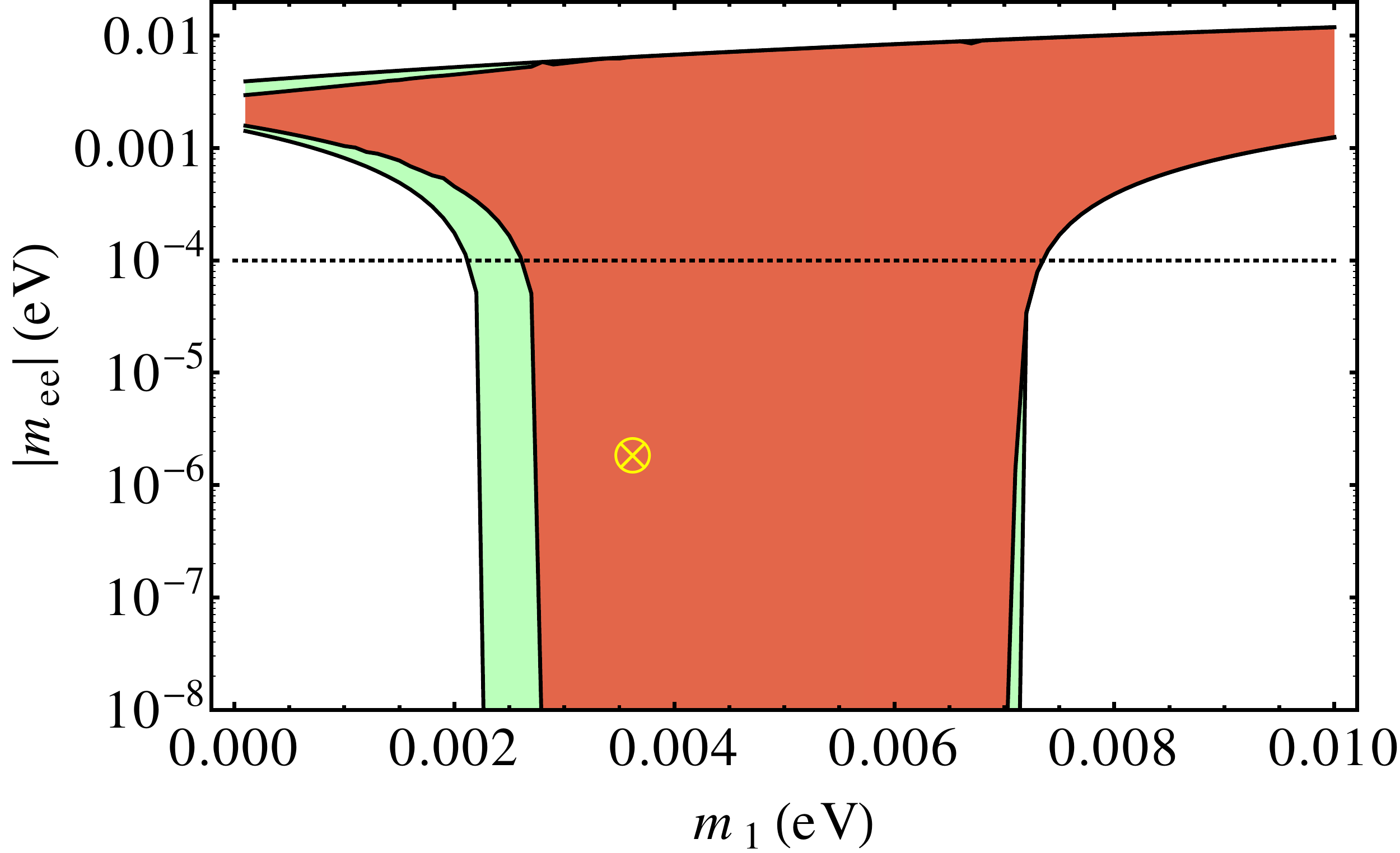}
\par\end{centering}
\caption{$|(m_{\nu})_{ee}|$ as a function of $m_1$. 
Similarly to Fig.~\ref{fig:s13-mem}, but with the red,~darker region instead satisfying the extra constraint
 $|(m_{\nu})_{e\mu}| < |(m_{\nu})_{ee}|$. 
We also draw the line $|(m_{\nu})_{ee}| = 10^{-4}$, which is the 
upper-limit estimate in this model.
\label{fig:0nubetabeta-m1}}
\end{figure}

Of course, $(m_{\nu})_{ee}$ and $(m_{\nu})_{e\mu}$ cannot be identically zero 
but small, $\lesssim 10^{-4} $ eV. In fact,
$g_{ee}$ must be different
from zero and rather large in order to $0\nu\beta\beta$ decay be 
observable (in this type of models  $(m_{\nu})_{ee}$ is small 
due to the huge suppression factor $m_{e}^{2}$ 
entering in its expression, not because $g_{ee}$ is small itself). 
When $(m_{\nu})_{ee}$ and $(m_{\nu})_{e\mu}$ are allowed to vary within the model, 
with the other parameters staying within their  1~$\sigma$ range, $\sin^2\theta_{13}$ is no 
longer bounded  from above although the lower bound remains:
\begin{equation}
\sin^2\theta_{13} \gtrsim 0.008\ .
\end{equation}
In order to illustrate this behaviour we plot in Fig. \ref{fig:s13-mem} 
$|(m_{\nu})_{e\mu}|$ as a function of 
$\sin^2\theta_{13}$, with all the neutrino masses and other mixing 
parameters varying arbitrarily within their 1~$\sigma$ limits. 
The red, darker region corresponds
to $|(m_{\nu})_{ee}|$ less than $|(m_{\nu})_{e\mu}|$. 
For comparison, we also plot the recent Double Chooz limit given above which is fully 
compatible with the 
T2K 90\% C.L. interval $0.007 - 0.07$ \cite{Abe:2011sj}, 
and in agreement with current global fits (for instance, the one used in this paper 
\cite{Schwetz:2011zk} 
allows for $0.008<\sin^2\theta_{13} < 0.020$  at $1~\sigma$).
Analogously, in Fig.~\ref{fig:0nubetabeta-m1} we draw $|(m_{\nu})_{ee}|$ as a function of 
the lightest neutrino mass  $m_1$.
The red, darker region stands now for $|(m_{\nu})_{e\mu}| < |(m_{\nu})_{ee}|$. 
The previous figures are obtained from the neutrino mass restrictions only. 
But in our approach the model parameters are further constrained 
by the bounds from $0\nu\beta\beta$ decay and LFV processes, which, as already emphasized, require
 $|(m_{\nu})_{ee, e\mu}| \lesssim 10^{-4}$ 
(see Eqs. (\ref{eq:mee-pert-bound}) and (\ref{eq:mem.bound}), respectively). 
Then, 
$\sin^2\theta_{13} \gtrsim 0.008$ and $0.002\ {\rm eV} \lesssim m_1 \lesssim 0.007\ {\rm eV}$, 
as seen from Figs. \ref{fig:s13-mem} and \ref{fig:0nubetabeta-m1}, respectively.
It then follows that a sufficiently precise measurement of $\sin\theta_{13}$ can 
exclude the model being discussed, as stressed before. 

To conclude this section we derive
the lower bounds, announced in Section 
\ref{sec:0nu2beta}, on the 
scalar masses implied by the experimental limit on the 
$\tau^{-}\rightarrow e^{+}\mu^{-}\mu^{-}$ branching ratio 
($<1.7\times 10^{-8}$). 
In our model $|g_{\tau e}g_{\mu\mu}^*|$ must be large 
in order to reproduce the observed pattern of neutrino masses. 
More precisely, 
\begin{equation}
|g_{e\tau}g_{\mu\mu}|=\frac{\left(2(2\pi)^{4}v_{\phi}^{4}\right)^{2}}{m_{e}m_{\tau}m_{\mu}^{2}}
\frac{|\left(m_{\nu}\right)_{e\tau}||\left(m_{\nu}\right)_{\mu\mu}|}
{\mu_{\kappa}^{2}v_{\chi}^{4}I_{\nu}^{2}}\stackrel{\mathrm{{\scriptstyle Pert}}}{>} 
0.065\left(\frac{\mathrm{TeV}}{I_\nu{\rm min}({m}_{\kappa_{1,2}})}\right)^{2} \ ,
\label{eq:taudecay-limit}
\end{equation}
where the second inequality follows from Eq.~\eqref{eq:pulimits} and the 
substitution of the other variables by their approximate values (see, for 
instance, Eq.~\eqref{eq:mnu-example} for the $|\left(m_{\nu}\right)_{e\tau,\mu\mu}|$ 
estimates), in particular $v_{\chi} < 2$ GeV. 
Then, using Eq.~\eqref{eq:Rmuto3e} for $\tau^{-}\rightarrow e^{+}\mu^{-}\mu^{-}$ 
and the experimental limit on its branching ratio 
one obtains $|g_{e\tau}g_{\mu\mu}|<0.007 (m_\kappa/\mathrm{TeV})^2$,  which combined with 
Eq.~\eqref{eq:taudecay-limit} yields $m_{\kappa}>1.2\,\mathrm{TeV}$ 
(see Fig.~\ref{fig:plot-mk-mx}, left)~\footnote{Note that we have used $I_\nu \sim 2$ because 
$m_\chi$ is somewhat larger than $m_\kappa$ (see Appendix~\ref{sec:loop-integrals}).}. 
We will present a detailed study of LFV processes in this type of models elsewhere.   

\section{Collider signals\label{sec:colliders}}

Direct evidence for this type of models would be the 
discovery of the new scalars at a large collider. 
Doubly-charged scalars have fixed couplings to photons and 
are then produced at colliders with known cross sections. 
In addition their decay into leptons offers a very clean signal, 
which is particularly important at hadronic machines. 
Therefore, if doubly-charged scalars are light enough, they are
very well suited for detection at colliders. 

In general, this type of scalars is assumed to be part of a weak triplet, 
and usually also acts as see-saw messenger of type II 
generating tree-level Majorana masses for the 
light neutrinos \cite{Konetschny:1977bn,Cheng:1980qt,Schechter:1980gr,Gelmini:1980re} 
(see also \cite{Ma:1998dx}).
Such triplets are then well-motivated on theoretical grounds, especially
when considering LR symmetric models, and
studies for searching the corresponding 
doubly-charged scalars at future colliders 
have been performed in the past 
\cite{Gunion:1989in,Huitu:1996su,Gunion:1996pq,Akeroyd:2005gt,Azuelos:2005uc}
(see also \cite{delAguila:2008cj,Akeroyd:2010ip} for recent studies;
model independent studies have  also been carried out in 
the literature \cite{Dion:1998pw,Cuypers:1996ia}). 
The general conclusion is that the LHC discovery limit reaches masses
over $600$ GeV (for a center of mass energy of $14$ TeV and an integrated 
luminosity of 30 fb$^{-1}$) \cite{delAguila:2008cj,delAguila:2009bb}. 
However, the actual limits may be much better given 
the outstanding LHC performance, which almost matches the 
most favourable expectations for a CM energy of $7$~TeV~\cite{delAguila:2010uw}.
(See for a review \cite{Nath:2010zj}.) 

Recently, first results from CMS have been presented at a CM energy of 
$7\,$TeV with an integrated 
luminosity of $0.89\,\mathrm{fb^{-1}}$~\cite{cms-pas-hig-11-007:2011}.
The analysis  assumed a scalar triplet coupled
to leptons, with $100 \%$ branching ratio to each leptonic channel. 
Nothing is seen, leading to a lower bound on the doubly-charged scalar mass 
of about $250\,\mathrm{GeV}$ if the main decay
channel contain $\tau$ leptons, and to about $300\,\mathrm{GeV}$ 
if they contain only electrons or muons. 
Weaker limits were obtained previously by LEP and the Tevatron.
The absence at LEP of a pair-production signal
($e^{+}e^{-}\rightarrow\gamma^{\ast}, Z^{\ast}\rightarrow\kappa\bar{\kappa}$)
gives the constraint $m_{\kappa}>100\,\mbox{GeV}$ 
\cite{Abdallah:2002qj,Abbiendi:2001cr,Achard:2003mv}~\footnote{
Single production via $e^{+}e^{-}\rightarrow\kappa ee$, as well the
$u$-channel contribution of $\kappa $
to Bhabha scattering have also been studied 
at LEP \cite{Achard:2003mv,Abbiendi:2003pr}, but
the corresponding bounds depend on the unknown values of the
Yukawa couplings.}. Limits on this
type of scalars have been also derived using Tevatron data
\cite{Abazov:2004au,Acosta:2004uj,Acosta:2005np}, leading to a
limit $m_\kappa > 100-150\,\mathrm{GeV}$  (depending on the details of the model).

In our model the triplet does not directly couple to fermions, while the doubly-charged 
singlet does not couple to $W$ pairs; however, triplet and singlet mix. Of the resulting 
mass eigenstates one, $\kappa_{1}$, is mainly a singlet and decays dominantly to lepton pairs, 
while the other, $\kappa_{2}$, is mainly a triplet and has suppressed couplings to charged 
leptons. Both of them can be produced at LHC via the Drell-Yan mechanism 
($q\bar{q}\rightarrow\gamma^*, Z^*\rightarrow\kappa^{++}\kappa^{--}$)
with full strength. Since this  is the main production process considered
by CMS, the former limits apply directly  to $\kappa_{1}$ for a non-negligible mixing:
$300\,{\mathrm {GeV}} < m_{\kappa_{1}}$. Limits on $ m_{\kappa_2}$
will be more difficult to obtain since the process  
$q\bar{q}\rightarrow \gamma^*, Z^*\rightarrow {\kappa}_2\bar{\kappa}_2\rightarrow W^+W^+W^-W^-$ 
is much more complicated to deal with, due to its large backgrounds and the 
in general difficult reconstruction of several leptonic $W$ decays. 

Notice that there are other production processes 
that are more model
dependent. In particular, the same interaction that induces $0\nu\beta\beta$ 
decay and the decay of $\kappa$ into gauge bosons can mediate single-scalar 
production through $WW$ fusion.  
The amplitude is proportional to the triplet VEV, $v_{\chi}$, which
is small, and to the singlet-triplet mixing, $\sin\theta_{D}$, 
further suppressing this process. 
Nonetheless this could prove to be the
dominant production channel at LHC~\cite{Huitu:1996su}
if $v_{\chi}>1\,\mathrm{GeV}$ and 
$m_{\kappa_{1,2}}>500\,\mathrm{GeV}$. 
This is especially relevant for our model
since both LHC and low-energy constraints require a 
relatively large $v_{\chi}$ as well as large scalar masses
(see Fig. \ref{fig:plot-mk-mx}), all of which is
in marked contrast to the constraints on 
triplet models with tree-level  type II see-saw
neutrino masses. A thorough study
of the various possibilities is somewhat involved 
\cite{Melfo:2011nx} and lies outside the scope of this
paper; we will revisit these aspects of our model in a
future publication.

\section{Conclusions \label{sec:conclusions}}

In this paper we have presented a simple model with  
a large $0\nu\beta\beta$ decay rate into RH electrons
through the exchange of the SM $W$ boson 
and new heavy scalars $\kappa, \chi$.
In this model the light 
neutrino masses,  
$0\nu\beta\beta$ decay and
LFV processes have a common origin, which provides
a simple description of these processes and also leads to a rather
constrained parameter space.
A $0\nu\beta\beta$ decay final state with RH electrons also occurs 
in left-right (LR) models (see \cite{Tello:2010am} for a detailed discussion), but 
generated by the exchange of
new heavy neutrinos $N$ and gauge bosons $W_R$. 
In these models this process is a priori decoupled from the 
light neutrino mass generation, and hence the rate for $0\nu\beta\beta$ decay can be 
large and at the same time the effective electron mass 
$(m_\nu)_{ee}$ small. Despite these differences 
both types of models have similar low-energy 
(below the electroweak scale) limits or, equivalently,
they represent very different UV completions 
of two similar effective Lagrangians
at scales below $ m_W $. A general discussion of the different 
alternatives using an effective Lagrangian approach is presented in a 
companion paper~\cite{delAguila:2011zz}.

The specific model we have discussed is one of a wide class of theories
with similar phenomenology. For example,
we choose to break LN explicitly by introducing one 
neutral scalar singlet ($\sigma$), but, as discussed in
Section \ref{sec:The-model} we could promote $\sigma $
to a complex singlet and insure the Lagrangian is LN invariant,
and this symmetry spontaneously broken; while the neutrino phenomenology remains similar,
it is simpler to discuss in our case. In general, this class of 
models require several new scalar multiplets 
to allow for non-vanishing scalar couplings to RH electrons, 
together with a chain of scalar couplings connecting them to the standard 
gauge boson triplet, and if desired, to the SM Higgs doublet 
in order to relate the new LN acting on the RH leptons 
to the ordinary LN associated with LH leptons. 
In such cases $0\nu\beta\beta$ decay into RH electrons 
and the light neutrino Majorana masses are again related although at different loop order,
as in our model. 

This model, which we considered in some detail, is based on an extended 
scalar sector containing three new multiplets: two isosinglets  and one 
isotriplet, the physical spectrum contains two 
light scalars (with masses $\sim v_{\phi,\sigma}$) and
5 heavy scalars (with masses $O($TeV$)$):
2 doubly-charged, 1 singly-charged and 2 neutral.
After writing the corresponding scalar potential and showing that there 
is a region of parameter space allowing for a local minimum along 
the desired direction, we have elaborated on their low-energy 
phenomenological implications. In deriving our quantitative predictions
we required for the model to remain
perturbative up to several tens of TeV.

As repeatedly stressed, the new scalars
mediate $0\nu\beta\beta$ decay into RH electrons, LFV processes 
and, at two loops, generate Majorana neutrino masses. 
By requiring $0\nu\beta\beta$ decay to be large enough to be 
observable at the next round of experiments, we have derived lower bounds on 
the coupling to RH electrons $g_{ee}$ and upper bounds on the masses of the 
exchanged scalars. On the other hand, the stringent 
experimental limits on LFV transitions 
(particularly $\mu^{-}\rightarrow e^{+}e^{-}e^{-}$ and 
$\tau^{-}\rightarrow e^{+}\mu^{-}\mu^{-}$) translate into upper bounds on 
the couplings $g_{ab}$ and lower bounds on the masses of the new particles involved, 
and these create some tension with the assumed $0\nu\beta\beta$ decay 
rate. As a result, the model predicts that
most LFV processes can be within the reach of the next generation 
of experiments, too.  

These same couplings and masses also enter the (two-loop) expression for the Majorana neutrino 
masses that have the very characteristic form, 
$(m_\nu)_{ab} \propto m_{a}g^*_{ab}m_{b}$, 
proportional to the scalar couplings to two RH leptons $g_{ab}$ and to the 
corresponding charged lepton masses $m_{a,b}$.
Accommodating the observed neutrino mass spectrum and mixing parameters 
together with the other constraints is possible, but only for
very restricted values of the lightest neutrino mass (not testable 
in neutrino oscillation experiments), and when the mixing 
angle $\theta_{13}$ is constrained to lie within the 
preferred 1~$\sigma$ range from the global fit in Ref.~\cite{Schwetz:2011zk}, 
$0.008 < \sin^2\theta_{13} < 0.020$.  
This last prediction also goes far beyond our model, and holds 
whenever the (symmetric) neutrino mass matrix 
contains three very small entries, for example, 
$m_{ee}, m_{e\mu} (= m_{\mu e}) \sim 0$, as in our case 
\footnote{A general discussion of the implications of texture 
zeroes in neutrino mass matrices can be found in 
\cite{Frampton:2002yf,Xing:2002ta}, and most recently in \cite{Fritzsch:2011qv}.}.  
Some of these constraints can be alleviated by further
extending the scalar sector, although at the price of requiring
precise cancellations (that could be naturally enforced using 
further symmetries).

The model considered presents a simple consistent extension of the
SM exhibiting a large $0\nu\beta\beta$ decay, but where the contributions 
to this decay generated by
the neutrino Majorana masses are negligible. The
neutrino masses themselves are predictable, in contrast with
other proposals (see, for example \cite{Brahmachari:2002xc}), and
consistent with existing data. 
Note, however, that although our phenomenological approach and 
the constraints on the model mainly follow from requiring an 
observable $0\nu\beta\beta$ in the next round of experiments, 
the non-observation of this decay and even a vanishing decay rate 
would be compatible with our analysis. That would be the case for  
$g_{ee}\rightarrow 0$, value which would also ease the LFV restrictions 
without altering the neutrino mass predictions. 
As they are obtained assuming that $(m_\nu)_{ee}$, which is proportional 
to $g_{ee}$, too, is to a large extend negligible.

Finally, we have also reviewed the collider limits on doubly-charged scalars.
The excellent LHC performance should soon allow for the actual confrontation
with the expected masses and couplings for the new scalars in the type of
models studied here. We must, anyhow, be aware
that once LHC settles the fate of the SM Higgs 
boson\footnote{In the very near future the LHC will be able to exclude a 
heavy Higgs with a relatively large
significance \cite{atlas-conf-2011-157,cms-pas-hig-11-023} but electroweak 
precision data do prefer a light Higgs 
(see, for instance, Ref.~\cite{delAguila:2011zs} and references there in), 
which shall require further for confirmation (or exclusion).}, 
the mass and couplings of the scalar doublet will be further 
constrained, implying further restrictions in the scalar potential which 
must be checked that can be satisfied. 
At any rate, the model studied here is one of a wider class sharing the 
main assumption in our analysis, that $0\nu\beta\beta$ 
decay is large and decoupled from any mechanism 
providing tree-level neutrino masses, although the latter 
are generated through higher-order radiative corrections.
  
\appendix

\section{Bound on the scalar triplet VEV \label{TripletVEV}}

As it is well known, the VEV of a triplet with hypercharge $1$ gives 
a tree-level contribution to the $\rho$ parameter, spoiling the successful
SM (tree-level) prediction $\rho_{0}=1$. 
In general, 
\begin{eqnarray}
\rho_{0}=\frac{\Sigma_{i}v_{i}^{2}[T_{i}(T_{i}+1)-Y_{i}^{2}]}{\Sigma_{i}2v_{i}^{2}Y_{i}^{2}} \ , 
\end{eqnarray}
where the sum runs over the scalars of isospin $T_{i}$ and hypercharge $Y_{i}$ with VEV $v_{i}$. 
For the case under consideration, and assuming that the VEV of the triplet is much 
smaller than the one of the doublet,  
\begin{eqnarray}
\rho_{0}=\frac{v_{\phi}^{2}+2v_{\chi}^{2}}{v_{\phi}^{2}+4v_{\chi}^{2}}\approx1-
\frac{2v_{\chi}^{2}}{v_{\phi}^{2}}\,.
\label{rho0}
\end{eqnarray}
Then, the VEV of the scalar triplet contributes negatively to $\rho_{0}$, 
while the best $\rho$ value obtained from a global fit to electroweak precision data is 
~\cite{Nakamura:2010zzi} $\rho=1.0008{+0.0017\atop -0.0010}$
at $1\sigma$, and $\rho=1.0004{+0.0023\atop -0.0011}$ at $2\sigma$. 
Thus, $v_{\chi}^{2}/v_{\phi}^{2}<0.00035$ at 2~$\sigma$, 
implying $v_{\chi}<3\,\mathrm{GeV}$ for $v_{\phi} \approx 174\,\mathrm{GeV}$. 
Which is comparable to the bound derived from the global fit including explicitly 
the scalar triplet effects, $v_{\chi}<2\,\mathrm{GeV}$ at the 90\% C.L. \cite{delAguila:2008ks}. 

However, a more complete analysis should also include the radiative corrections 
to the $\rho$ parameter induced, for example, by the exchange of the scalar triplet, which can be 
positive~\cite{Einhorn:1981cy}. For instance, in the triplet Majoron model
~\cite{Golden:1986jn} $\Delta\rho=\dfrac{(1-\ln2)}{2\pi^{2}\sqrt{2}}G_{F}m_{\chi}^{\text{2}}$, 
with $m_{\chi}$ the mass of the doubly-charged scalar, cancelling partially the 
tree-level contribution. 
Since these contributions depend on the mass splitting of the triplet components, 
which in our model is not fixed, we will assume a conservative upper bound 
\begin{equation}
v_{\chi}<5\,\mathrm{GeV}\,.
\label{eq:bound-triplet-vev}
\end{equation}

\section{Constraints from naturality and perturbative unitarity \label{Perturbativityconstraints}}

The relevant parameter for $0\nu\beta\beta$ decay and neutrino masses is the
product of couplings and VEVs $\mu_{\kappa} v_{\chi}^{2} g_{ee}$. 
It cannot be too large without leaving the perturbative regime and there
are different arguments that can be used to set upper limits
on its size (perturbative unitarity, naturality, etc.). 

Let us discuss the constraints from perturbative unitarity. 
Consider first the Yukawa couplings of doubly-charged scalars. 
Tree-level unitarity at high energy, $s\gg m_{\kappa_{1,2}}$, 
in $ee\rightarrow ee$ collisions mediated by $\kappa_{1,2}$ 
requires $|g_{ee}|<\sqrt{4\pi}$.
Similar bounds can be obtained from other channels. In order to be definite, 
we will demand 
\begin{equation}
|g_{\alpha\beta}|\stackrel{\mathrm{{\scriptstyle Pert}}}{<}\sqrt{4\pi} \ .
\label{eq:gs-ulimit}
\end{equation}
Tree-level unitarity at high energy does not give useful information on dimensional 
parameters like $\mu_{\kappa}$ because amplitudes involving these couplings 
decrease with energy. However, it does not seem natural to have $\mu_{\kappa}$ 
much larger than other dimensionful parameters in the model,
like $m_\kappa$ or $m_\chi$. In fact, one-loop diagrams involving the $\mu_\kappa$ 
coupling give contributions to $m_\kappa$ or $m_\chi$ which are of order 
$\delta m_{\kappa,\chi} \sim \mu^2_\kappa/(4\pi)^2$. Therefore, it seems appropriate to require   
\begin{equation}
\mu_{\kappa}\stackrel{\mathrm{{\scriptstyle Pert}}}{<}
4\pi\ {\rm min}(m_{\kappa_{1,2}})\,.
\label{eq:muk-ulimit}
\end{equation}
Limits \eqref{eq:gs-ulimit} and \eqref{eq:muk-ulimit} also guarantee that the 
$\kappa_1$ 
decay width (to leptons and to $\kappa_2$, if this is light enough) is not too 
large as compared to its mass.

One must be aware, however, that all these limits are estimates 
which depend on the naturality approach. 
Thus, although at the price of fine tuning, one might
decide to fix the model parameters outside the range defined by these limits. 
Moreover, there can be model extensions where those values are natural.  At any rate, 
we use Eqs. (\ref{eq:gs-ulimit}) and (\ref{eq:muk-ulimit}) in the text 
to illustrate that the allowed regions in parameter space 
are at a large extent bounded if the perturbative theory must stay 
natural.

\section{Loop integrals for evaluating neutrino masses
\label{sec:loop-integrals}}

\begin{figure}
\includegraphics[width=0.49\columnwidth]{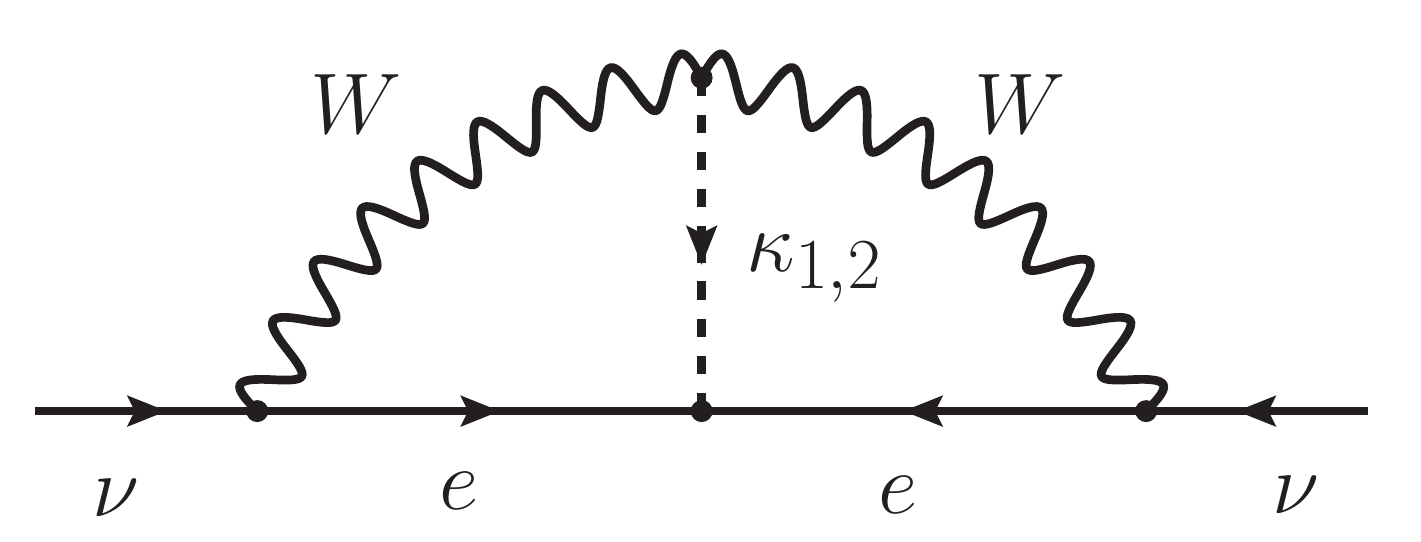}
\includegraphics[width=0.49\columnwidth]{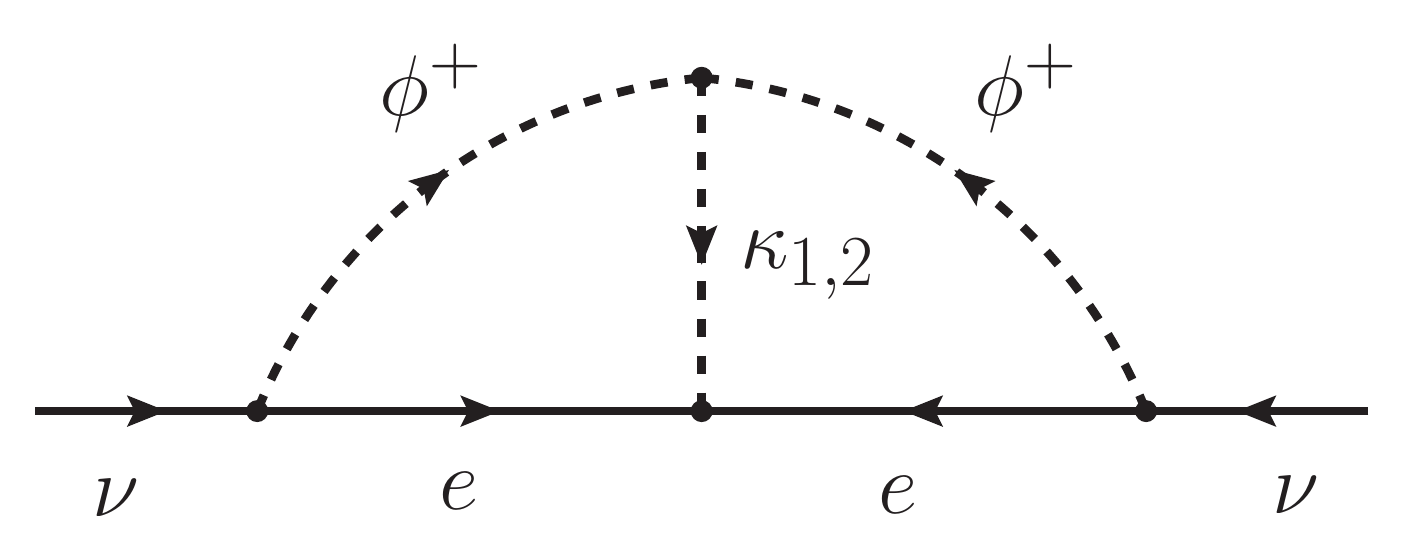}
\caption{Two-loop diagrams contributing to neutrino masses in the generalized $R_{\xi}$ 
gauge. 
\label{fig:numass-w-gauge}}
\end{figure}

The evaluation of the complete two-loop contribution to neutrino masses, 
including $m_{W}$ effects in a general $R_{\xi}$ gauge, is a complicated task. 
Fortunately, we found a gauge in which the calculation simplifies enormously. 
Indeed, one can choose a gauge in which triplet and doublet charged scalars
do not mix at all (the gauge-fixing Lagrangian is of the form
$ {\cal L}_{\rm gf}= 
a | \partial\cdot W^+ + b \chi^+ + c \phi^+ |^2 $ with the constants 
$a,b,c$ chosen to cancel the $ \phi^+ \chi^- $
and the $ W^+ \chi^-,~ W^+ \phi^- $ mixing terms). 
In this gauge the charged Goldstone
boson becomes degenerate with the physical charged scalar; and $\omega^{\pm}$
does not couple to fermions, whereas $G^{\pm}$ does not have derivative
couplings to the doubly-charged scalars and gauge bosons. Therefore, the
only diagrams contributing to neutrino masses are those depicted
in Fig.~\ref{fig:numass-w-gauge}. (The corresponding Feynman rules are obtained 
redefining the fields in Section ~\ref{sec:scalar-spectrum} adequately; in particular, in this gauge 
$\omega^\pm = \chi^\pm$ and $G^\pm = \phi^\pm$.)
Following the notation in Eq.~\eqref{eq:nu-mass-matrix} we will have
now $I_{\nu}=I_{W}+I_{\phi}$, where $I_{W}$ and $I_{\phi}$ are the
two contributions from the diagrams in Fig.~\ref{fig:numass-w-gauge}.  
They read 
\begin{equation}
I_{W}=-2(4\pi)^{4}m_{W}^{4}\cos^{4}\theta_{S}\int \frac{1}{k^{2}
(k^{2}-m_{W}^{2})q^{2}(q^{2}-m_{W}^{2})((k-q)^{2}-m_{\kappa_{1}}^{2})((k-q)^{2}
-m_{\kappa_{2}}^{2})}\times \nonumber
\end{equation}
\begin{equation}
\times\left(4-\left(1-\frac{m_{\omega}^{2}}{m_{W}^{2}}\right)\left(\frac{k^{2}}{k^{2}
-m_{\omega}^{2}}+\frac{q^{2}}{q^{2}-m_{\omega}^{2}}\right)+
\left(1-\frac{m_{\omega}^{2}}{m_{W}^{2}}\right)^{2}\frac{\left(k\cdot q\right)^{2}}
{(k^{2}-m_{\omega}^{2})(q^{2}-m_{\omega}^{2})}\right) \ , 
\end{equation}
\begin{equation}
I_{\phi}=(4\pi)^{4}\frac{m_{A}^{2}}{\cos^{2}\theta_{I}}\int \frac{k\cdot q}{k^{2}(k^{2}
-m_{\omega}^{2})q^{2}(q^{2}-m_{\omega}^{2})((k-q)^{2}-m_{\kappa_{1}}^{2})((k-q)^{2}
-m_{\kappa_{2}}^{2})} \ , \nonumber
\end{equation}
where in $I_{\phi}$ we used the equality 
$ v_\chi m^2_A = -\lambda_6 v_\sigma v^2_\phi \cos^2{\theta_I}$ 
(see Section \ref{sec:The-model}) to rewrite $\lambda_{6}$ in terms of $m_{A}$. 
Note also that both doubly-charged scalar masses must enter symmetrically 
in the integrals to obtain a non-vanishing contribution. 

\begin{figure}
\begin{centering}
\includegraphics[width=0.6\columnwidth]{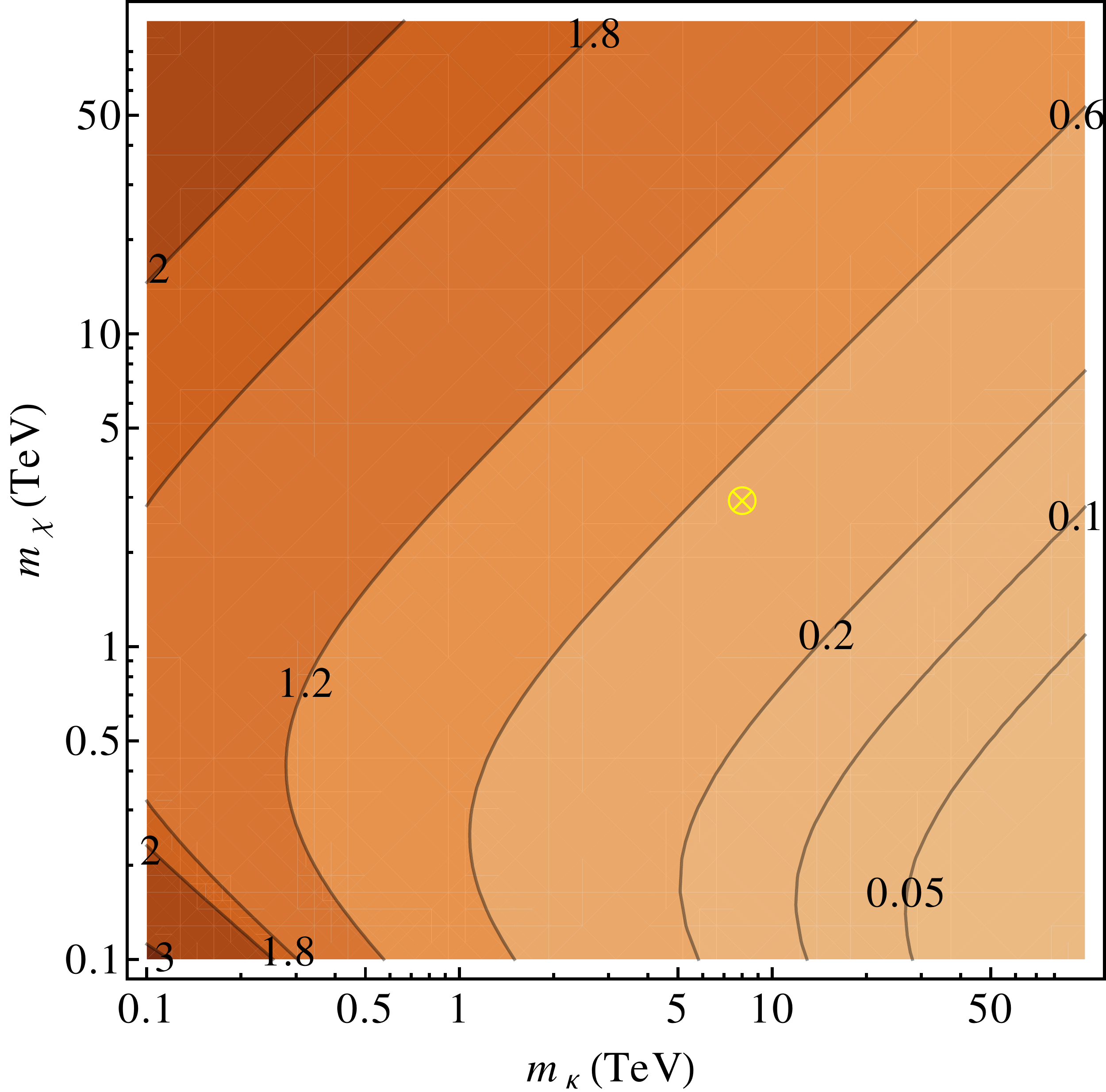}
\par\end{centering}
\caption{Contour plot for the loop integral $I_{\nu}$ as a function of the
masses $m_{\kappa_{1}} \approx m_{\kappa}$ and $m_{\kappa_{2}} \approx m_{\chi}$ 
for $m_{W}=80\,\mathrm{GeV}$ and $m_{\omega}=m_{A}=m_{\kappa_{2}}$. 
The cross denotes the reference point in the text. 
\label{fig:integral-cplot}}
\end{figure}
We have checked that in the decoupling limit 
($m_{W}\ll m_{\kappa_{1}, {\kappa_{2}}, {\omega}, {A}}$ and 
$v_{\chi}\ll v_{\phi}$, implying $m_{\kappa_{2}}=m_{\omega}=m_{\chi}$ and
$\cos\theta_{S}=\cos\theta_{I}=1$) we recover the results obtained 
in the mass insertion approximation discussed
in Section~\ref{sec:the neutrino mass}. 
In general $I_{\nu}$ is a complicated function of 
$m_{W}, m_{\kappa_{1}, {\kappa_{2}}, {\omega}, {A}}$ 
and the ratio $v_{\chi}/v_{\phi}$, which defines $\cos\theta_{S}$
and $\cos\theta_{I}$. However,  since $v_{\chi}\ll v_{\phi}$, we can safely
neglect the corresponding corrections 
(in the limit of large $m_{\chi}$ the triplet VEV $v_\chi$ is small,
and so is then the mixing of the triplet components with doublets and singlets; 
analogously, the contribution of doublet and singlet VEVs to the 
masses of the triplet components can be ignored). In 
this case $I_{\nu}$ becomes a function
only of $m_{\kappa_{1}}^{2}/m_{W}^{2}$ and $m_{\kappa_{2}}^{2}/m_{W}^{2}$ 
that we compute numerically.
In Fig.~\eqref{fig:integral-cplot} we plot the contours of constant 
$I_{\nu}$ as a function of the masses $m_{\kappa_{1}} \approx m_{\kappa}$ 
and $m_{\kappa_{2}} \approx m_{\chi}$, 
where we have fixed $m_{W}=80\,\mathrm{GeV}$. From the Figure we see
that $I_{\nu}$ is order one for a large region of the parameter space. 
Only when $m_{\kappa_{1}}\gg m_{\kappa_{2}}$ there is some suppression. 
The cross corresponds to the value of the reference point in Section \ref{sec:0nu2beta}.

\section*{Acknowledgments}

This work has been supported in part by the Ministry of Science and
Innovation (MICINN) Spain, under the grant numbers FPA2006-05294, 
FPA2008-03373, FPA2010-17915 and FPA2011-23897, by the Junta de Andaluc{\'\i}a 
grants FQM 101, FQM 03048 and FQM 6552, by the 
``Generalitat Valenciana'' grant PROMETEO/2009/128 and by the 
U.S. Department of Energy grant No.~DE-FG03-94ER40837. 
A.A. is supported by the MICINN under the FPU program. 


\providecommand{\href}[2]{#2}\begingroup\raggedright\endgroup

\end{document}